\documentclass[english]{article}
\usepackage[latin9]{inputenc}

\usepackage{textcomp}
\usepackage[colorlinks]{hyperref}

\usepackage{amstext}
\usepackage{graphicx}
\usepackage{babel}
\usepackage{color}
\usepackage{tikz}
\usepackage{pict2e}
\usepackage{pgfplots}
\usepackage{float}

\restylefloat{table}

\font\tenbb=msbm10
\font\sevenbb=msbm7
\font\fivebb=msbm5
\newfam\bbfam
\textfont\bbfam=\tenbb
\scriptfont\bbfam=\sevenbb
\scriptscriptfont\bbfam=\fivebb

\hypersetup{
    citecolor=blue,
    linkcolor=blue,
    filecolor=blue,
    urlcolor=blue,% color of links to bibliograph        % color of external links
}

\newcommand{\freq}{\mathrm{freq}}
\newcommand{\conf}{\mathrm{conf}}
\newcommand{\supp}{\mathrm{supp}}

\newcommand{\ARMTrav}{\mathrm{ARM\mbox{-}Traversal}}
\newcommand{\FIMTrav}{\mathrm{FIM\mbox{-}Traversal}}
\newcommand{\Rule}{\mathrm{RULE}}

\newcommand{\DEL}{\mathrm{DEL}}

\begin{document}

\newtheorem{theo}{Theorem}
\newtheorem{prop}{Proposition}
\newtheorem{property}{Property}
\newtheorem{lemma}{Lemma}
\newtheorem{cor}{Corollary}
\newtheorem{defi}{Definition}
\newtheorem{rque}{Remark}

\title{A Recursive Algorithm for Mining Association Rules
}
\author{Abdelkader Mokkadem\footnote{Universit\'e Paris-Saclay, UVSQ, CNRS, Laboratoire de Math\'ematiques de Versailles, 78000, Versailles, France (abdelkader.mokkadem@uvsq.fr)} \& 
Mariane Pelletier\footnote{Universit\'e Paris-Saclay, UVSQ, CNRS, Laboratoire de Math\'ematiques de Versailles, 78000, Versailles, France (mariane.pelletier@uvsq.fr)} \& Louis Raimbault\footnote{Universit\'e Paris-Saclay, UVSQ, CNRS, Laboratoire de Math\'ematiques de Versailles, 78000, Versailles, France and MAISONS DU MONDE FRANCE SAS, Le Portereau, 44120 Vertou, France (louis.raimbault@ens.uvsq.fr)}}
\date{}
\maketitle

\begin{abstract}
Mining frequent itemsets  and association rules is an essential task within data mining and data analysis. In  this paper, we introduce PrefRec, a recursive algorithm for finding frequent
itemsets and association rules. Its main advantage is its recursiveness with respect to the items. It is particularly efficient  for updating the mining process when new items are added to the database or when some are excluded. We present in a complete way the logic of the algorithm, and give some of its applications. After that, we carry out an experimental study on the effectiveness of PrefRec. We first compare the execution times with some very popular frequent itemset mining algorithms. Then, we do experiments to test the updating capabilities of our algorithm. 
\end{abstract}

%\tableofcontents{}


\section{Introduction}
\textcolor{black}{Frequent itemset and association rule mining is
one of the fundamental problems in data mining and computational statistics. This data analysis method} was first introduced
by Agrawal et al.~\cite{Agrawal1993} for mining transaction databases.\textcolor{black}{{}
Even if this method was introduced in the context of Market Business
Analysis, it has many applications in other fields, such as genetics,
webmining, textmining.} The problem can be stated as follows (Agrawal et al.~\cite{Agrawal1993}). Let $I=\{a_{1},\ldots,a_{n}\}$ be a finite set
of items. A transaction database is a set of transactions $T=\{t_{1},\ldots,t_{N}\}$
where each transaction $t_{i}\subset I$, $1\le i\le N$, represents
a nonempty subset of items. An itemset $A$ is a subset of $I$; $A$
is a $k$-itemset if it contains $k$ items. The support of an itemset
$A$ is denoted as $\supp(A)$ and is defined as the number of transactions
which contain $A$. The relative support of $A$ is $\freq(A)=\supp(A)/N$.
$A$ is frequent if $\freq(A)\ge\sigma$ where $\sigma$ is a user-specified
minimum relative support threshold, called minSup. \textcolor{black}{The
problem of Frequent Itemset Mining (FIM) is to find all the frequent
itemsets for any given minSup.}

\textcolor{black}{An association rule is an implication $A\Rightarrow B$
where $A$ and $B$ are two itemsets.} The support of a rule $A\Rightarrow B$
is defined as $\supp(A\Rightarrow B)=\supp(A\cup B)$, its relative support as  $\freq(A\Rightarrow B)=\supp(A\Rightarrow B)/N$. The confidence
of a rule $A\Rightarrow B$ is defined as $\conf(A\Rightarrow B)=\supp(A\Rightarrow B)/\supp(A)=\freq(A\cup B)/\freq(A)$.
\textcolor{black}{The problem of Association Rule Mining} (ARM)  is to find
all association rules  having a relative support no less
than minSup and a confidence no less than a user-defined threshold called minConf. \textcolor{black}{Thus the problem must be solved in two steps:}

\textbf{\textit{Step 1. Frequent itemset mining}} $(FIM)$ is to determine
all the frequent itemsets, that is,  all the itemsets having a relative support no
less than minSup.

\textbf{\textit{Step 2. Association rule mining}} $(ARM)$ is to discover all the 
association rules \textcolor{black}{satisfying the confidence condition}
by using the frequent itemsets found in Step~1.

It is well known that the first step is the key to discovering all the association
rules. \textcolor{black}{After determining the frequent itemsets in
the first step, the solution to the second step is quite straightforward: just generate, for each frequent itemset $C$,  all the rules $A\Rightarrow B$
where $A$ is a subset of $C$ and $B=C\setminus A$. }

To carry out the Step 1 (FIM), we have to organize all the itemsets. For that we use the order of the database $I$ from which we deduce the prefix tree; it is a tree whose nodes are the itemsets and in which the father-child relation is a relation of inclusion. Most known algorithms use this structure.

The first efficient algorithm to frequent itemset mining is Apriori
(Agrawal and Srikant~\cite{Agrawal1994},
Agrawal et al.~\cite{Agrawal1996}). Since the Apriori algorithm was proposed,
there have been extensive studies on the improvements or extensions
of Apriori, e.g., hashing technique (Park et al.~\cite{Park1995}), partitioning
technique (Savasere et al.~\cite{Savasere1995}), sampling approach (Toivonen~\cite{Toivonen1996}),
dynamic itemset counting (Brin et al.~\cite{Brin1997}), incremental mining (Cheung et al.~\cite{CheungWong1996}), parallel and distributed mining (Agrawal and Shafer~\cite{AgrawalShafer1996}; Cheung et al.~\cite{Cheung1996}; Zaki et al.~\cite{Zaki1997}), integrating
mining with relational database systems (Sarawagi et al.~\cite{Sarawagi1998}), information stored in bitMatrix (Huang et al.~\cite{Huang2000}). As review papers on frequent itemset mining, we can cite Han et al.~\cite{Han2007} and Luna et al.~\cite{Luna2019}.

Among the most efficient algorithms that have been proposed, the most  famous ones are Eclat (Zaki~\cite{Zaki2000}), Fp-Growth (Han et al.~\cite{Han2000}) and  LCM
(Uno et al.~\cite{Uno2003}).
Both Apriori and Fp-growth methods mine frequent patterns from a set
of transactions in horizontal data format, while Eclat algorithm explores
the vertical data format and LCM uses both vertical and horizontal representation.

In this paper we introduce PrefRec, a new algorithm for fast discovery
of frequent itemsets and association rules mining. This algorithm is based on a construction of the prefix tree in successive stages. More precisely, for each $k$, we construct the prefix tree of the base $\{a_{1},\ldots,a_{k+1}\}$ using $a_{k+1}$ and the  prefix tree of the base $\{a_{1},\ldots,a_{k}\}$. One main property of the algorithm is that it is recursive with
respect to the items. This means that once the algorithm has treated
the set of items $I=\{a_{1},\ldots,a_{n}\}$, if a new item $a_{n+1}$
is inserted into $I$, then updating  the algorithm only uses  the item
$a_{n+1}$ and the result obtained previously with $I=\{a_{1},\ldots,a_{n}\}$.
This property is particularly important in the case where the database is not fixed and the items arrive successively, $k$ denoting the order of arrival of the item $a_{k}$. An example is that of a binary time series observed on N individuals at times $k\in \{1,\ldots,n\}$.
Other types of update are also straightforward, such as removing the
first item, adding or removing a set of items.

We also carry out an experimental study. The first part of the experimental study is focused  on the time performance of the frequent itemset mining algorithms and use several datasets. We consider PrefRec and some known algorithms. The first one is Apriori, the most used in many areas. The others are Eclat, Fp-Growth and LCM, and are among the most efficient algorithms.
It emerges from this study that PrefRec supports the comparison very well and that it performs better in some cases. It also appears that PrefRec is particularly efficient in the case of time series. This is probably because it is recursive.
In the second part of the experimental study, we carry out experiments concerning recursion (adding or removing items) that show the nice properties of PrefRec in this field.\\ 
For PrefRec, we use our own C++ implementation available at \url {https://github.com/LouisRaimbault/PrefRec}.

The rest of the paper is organized as follows. In Section~\ref{section formalism},
we give some formalism and definitions to describe the data set of transactions,
the items and the itemsets. Then we present the recursive building
of the tree of all itemsets and that of all frequent itemsets; we
give the properties of these trees.
{In Section~\ref{section algo}, we present the two versions of our algorithm (the one for the FIM and the one for the ARM), their updating, and their applications.}
In Section~\ref{section comparaison}, we give our experimental study and compare
the performance of PrefRec with that of other algorithms.

\section{FIM, ARM and Trees}

\label{section formalism}

\subsection{Usual Definitions in Data Mining}

Let $I=\{a_{1},a_{2},\ldots,a_{n}\}$ be a finite set of items called
the item base. A transaction database is a set of transactions $T=
\{t_{1},t_{2},\ldots,t_{N}\}$, 
where each transaction $t_{i}\subset I$, $1\le i\le N$, is nonempty.  An itemset $A$ is a subset of $I$;
$A$ may be the empty set $\emptyset$. The itemset $A$ is a $k$-itemset
if it contains $k$ items. The support of an itemset $A$ is denoted
as $\supp(A)$ and is defined as the number of transactions $t_{i}$
such that $A\subset t_{i}$. The relative support of $A$ is $\freq(A)=\supp(A)/N$;
in the trivial case $A=\emptyset$, $\supp(A)=N$ and $\freq(A)=1$.
We say that $A$ is frequent if $\freq(A)\ge\sigma$ where $\sigma$ is
a user-specified minimum relative support threshold, called minSup.

A rule is an implication of the form $A\Rightarrow B$ where $A$
and $B$ are itemsets such $A\cap B=\emptyset$; $A$ is called the
antecedent and $B$ the consequent.  The support of a rule
$A\Rightarrow B$ is defined as $\supp(A\Rightarrow B)=\supp(A\cup B)$.
The confidence of a rule $A\Rightarrow B$ is defined as $\conf(A\Rightarrow B)=\supp(A\Rightarrow B)/\supp(A)=\freq(A\cup B)/\freq(A)$.
The problem of mining association rules is to find all the association
rules in a database having a relative support no less than minSup and a confidence no less than a user-defined minimum
confidence threshold called minConf.

Let $C$ be  an itemset; for each $A\subset C$ we denote by $R(A,C)$
the association rule $A\Rightarrow C\setminus A$. The support of
$R(A,C)$ is then $\supp(C)$ and the confidence $\supp(C)/\supp(A)$.
Given $C$, there are two trivial rules: $R(C,C)$ whose confidence
is $1$, and $R((\emptyset),C)$ whose confidence is $\freq(C)$. Note
that the rule $R((\emptyset),C)$ means ``C is in the transaction''.
Of course, these trivial rules are not to be considered. Let $\sigma$
and $\tau$ be  the minSup and the minConf, respectively. If $C$ is a
frequent itemset, i.e. an itemset such that $\freq(C)\ge\sigma$, then for each
$A\subset C$, $A$ is frequent. Let $\Rule(C,\tau)=\{R(A,C),A\subset C,\conf(R(A,C))\geq\tau\}$
be the set of association rules of the form $A\Rightarrow C\setminus A$
and satisfying the confidence condition. It is then easy to see that
the set of all association rules satisfying the support and the confidence conditions
is the union of the sets $\Rule(C,\tau)$ with $C$ frequent.

\subsection{Formal Interpretation of the Itemsets}

The item base $a_{j}$, $1\le j\le n$, is ordered by the index $j$. Thus $I=\{a_{1},a_{2},\ldots,a_{n}\}$ will  also be written $I=\{1,2,\ldots,n\}$. Likewise, each itemset $A=\{a_{j_{1}},\ldots,a_{j_{k}}\}$ will also be represented by a vector $(j_{1},j_{2},\ldots,j_{k})$, where the coordinates are increasing.
The database $T$ can therefore be represented as a sequence $\{t_{1},t_{2},\ldots,t_{N}\}$ where each $t_{i}$ is a vector with increasing coordinates in $\{1,2,\ldots,n\}$.

Another way to represent T is as follows.
 To each item $a_{j}$, $1\le j\le n$, we associate a
random variable $X^{(j)}=1$ if $a_{j}$ is in the transaction and
$X^{(j)}=0$ if not. Each transaction $t$ is thus seen as an observation
of the random variable $(X^{(1)},\ldots,X^{(n)})$. The database $T=\{t_{1},t_{2},\ldots,t_{N}\}$
is then a sample $(x_{i}^{(1)},\ldots,x_{i}^{(n)})_{1\leq i\leq N}$ of
the random variable $(X^{(1)},\ldots,X^{(n)})$. Equivalently, the database $T$ is the $N\times n$ matrix whose line~$i$ is $t_{i} = (x_{i}^{(1)},\ldots,x_{i}^{(n)})$; the  column $x^{(j)}$ represents the  item $a_{j}$.

To each $k$-itemset $A=\{a_{j_{1}},\ldots,a_{j_{k}}\}$, we associate the random variable
$Z^{(A)}=X^{(j_{1})}\times X^{(j_{2})}\times \ldots \times X^{(j_{k})}$, which clearly takes
the value $1$ if and only if the transaction contains $A$.  We note that  $(x_{i}^{(j_{1})}\times x_{i}^{(j_{2})}\times \ldots \times x_{i}^{(j_{k})})_{1\leq i\leq N}$ are the observed values of $Z^{(A)}$, so that $\freq(A)=\frac{1}{N}\sum_{i=1}^{N}x_{i}^{(j_{1})}\times x_{i}^{(j_{2})}\times \ldots \times x_{i}^{(j_{k})}$. 
In the particular case $A=(\emptyset)$, $Z^{(A)}$ is the constant variable
whose unique value is $1$, and $\freq(A)=1$.

\subsection{Tree Enumeration of the Itemsets}

\subsubsection{Trees and  Depth First Search}

The subset relationships between itemsets form a partial order on
the set itemsets. The most complete way to describe this set is the
Hasse diagram; it is the graph where the nodes are the itemsets and {{where}}
there is a an edge between the two itemsets $I$, $J$ if and only
if  $J=I\cup\{a\}$ where $a$ is an item. However,
this representation leads to redundant search because the same itemset   can be built multiple times by adding its items in different orders.
To eliminate this redundancy, the Hasse graph is usually reduced to a tree called prefix tree.

\begin{figure}[h!]

\begin{center}
% Racine en Haut, développement vers le bas
\tikzset{
    ultra thin/.style= {line width=0.1pt},
    very thin/.style=  {line width=0.2pt},
    thin/.style=       {line width=0.4pt},% thin is the default
    semithick/.style=  {line width=0.6pt},
    thick/.style=      {line width=0.6pt},
    very thick/.style= {line width=1.1pt},
    ultra thick/.style={line width=1.3pt}
}
\begin{tikzpicture}[xscale=1,yscale=1]
% Styles (MODIFIABLES)

\tikzstyle{noeud} = [rectangle,rounded corners ,fill = white,ultra thick,minimum height=5mm, minimum width = 5mm ,font=\bfseries\fontsize{8}{0}\selectfont,draw]

% Dimensions (MODIFIABLES)
\def\DistanceInterNiveaux{1.4}
\def\DistanceInterFeuilles{1.4}
% Dimensions calculées (NON MODIFIABLES)
\def\NiveauA{(-0)*\DistanceInterNiveaux}
\def\NiveauB{(-0.6)*\DistanceInterNiveaux}
\def\NiveauC{(-1.2)*\DistanceInterNiveaux}
\def\NiveauD{(-1.8)*\DistanceInterNiveaux}
\def\InterFeuilles{(1)*\DistanceInterFeuilles}
% Noeuds (MODIFIABLES : Styles et Coefficients d'InterFeuilles)
\node[noeud] (R) at ({(2)*\InterFeuilles},{\NiveauA}) {R};

\node[noeud] (B1) at ({(-0.5)*\InterFeuilles},{\NiveauB}) {};
\node [noeud](B2) at ({(1.35)*\InterFeuilles},{\NiveauB}) {};
\node [noeud](B3) at ({(2.65)*\InterFeuilles},{\NiveauB}) {};
\node [noeud](B4) at ({(4.5)*\InterFeuilles},{\NiveauB}) {};

\node [noeud](C1) at ({(-1.85)*\InterFeuilles},{\NiveauC}) {};
\node [noeud](C2) at ({(-0.75)*\InterFeuilles},{\NiveauC}) {};
\node [noeud](C3) at ({(0.15)*\InterFeuilles},{\NiveauC}) {};

\node [noeud](C4) at ({(1.20)*\InterFeuilles},{\NiveauC}) {};

\node [noeud](C5) at ({(2.30)*\InterFeuilles},{\NiveauC}) {};
\node [noeud](C6) at ({(3)*\InterFeuilles},{\NiveauC}) {};

\node [noeud](C7) at ({(4.25)*\InterFeuilles},{\NiveauC}) {};
\node [noeud](C8) at ({(5.00)*\InterFeuilles},{\NiveauC}) {};
\node [noeud](C9) at ({(5.85)*\InterFeuilles},{\NiveauC}) {};

\node [noeud](D1) at ({(3.7)*\InterFeuilles},{\NiveauD}) {};
\node [noeud](D2) at ({(4.65)*\InterFeuilles},{\NiveauD}) {};

% Arcs (MODIFIABLES : Styles)

\draw[-,line width = 0.4mm]  (R.south) to (B1.north);
\draw[-,line width = 0.4mm]  (R.south) to (B2.north);
\draw[-, line width = 0.4mm]  (R.south) to (B3.north);
\draw[-, line width = 0.4mm]  (R.south) to (B4.north);

\draw[-, line width = 0.4mm]  (B1.south) to (C1.north);
\draw[-, line width = 0.4mm]  (B1.south) to (C2.north);
\draw[-, line width = 0.4mm]  (B1.south) to (C3.north);

\draw[-, line width = 0.4mm]  (B2.south) to (C4.north);

\draw[-, line width = 0.4mm]  (B3.south) to (C5.north);
\draw[-, line width = 0.4mm]  (B3.south) to (C6.north);

\draw[-, line width = 0.4mm]  (B4.south) to (C7.north);
\draw[-, line width = 0.4mm]  (B4.south) to (C8.north);
\draw[-, line width = 0.4mm]  (B4.south) to (C9.north);

\draw[-, thick,line width = 0.4mm]  (C7.south) to (D1.north);
\draw[-, thick,line width = 0.4mm]  (C7.south) to (D2.north);

\end{tikzpicture}

\end{center}

 \caption{The tree T with 16 nodes and a root R}
\label{fig 1} 
\end{figure}
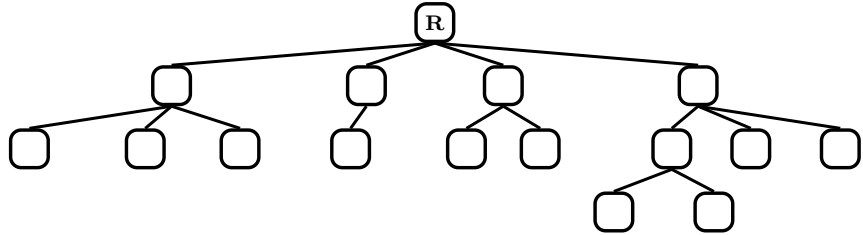

Let us briefly recall some elements on trees. We will only consider planar trees and thus we define a tree as a planar connected graph without cycle. We always assume the tree has a root, i.e. a
distinguished node. Figure~\ref{fig 1} gives a tree $T$ with 16
nodes and a root R.

%\medskip{}

\begin{figure}[h!]
\begin{center}
% Racine en Haut, développement vers le bas

\begin{tikzpicture}[xscale=1,yscale=1]
% Styles (MODIFIABLES)

\tikzstyle{bigpoint} = [fill = black, circle, draw, scale = 0.6]
\tikzstyle{texto} = [thick, scale = 1.3]
\tikzstyle{invisiblepoint} = [fill = black, circle, scale = 0.01]

% Dimensions (MODIFIABLES)
\def\DistanceInterNiveaux{1.4}
\def\DistanceInterFeuilles{1.4}
% Dimensions calculées (NON MODIFIABLES)
\def\NiveauA{(-0)*\DistanceInterNiveaux}
\def\NiveauABis{-0.4)*\DistanceInterNiveaux}
\def\NiveauB{(-0.8)*\DistanceInterNiveaux}
\def\NiveauC{(-1.3)*\DistanceInterNiveaux}
\def\NiveauD{(-1.8)*\DistanceInterNiveaux}
\def\InterFeuilles{(1)*\DistanceInterFeuilles}
% Noeuds (MODIFIABLES : Styles et Coefficients d'InterFeuilles)
\node[bigpoint] (R) at ({(3.15)*\InterFeuilles},{\NiveauA}) {};
\node [texto] (Rt) at ({(3.6)*\InterFeuilles},{\NiveauA}) {$Root$};
\node[invisiblepoint](Rbis) at ({(2.5)*\InterFeuilles},{\NiveauABis}){};
\node[bigpoint] (f) at ({(2)*\InterFeuilles},{\NiveauB}) {};
\node [texto] (tf) at ({(2.3)*\InterFeuilles},{\NiveauB}) {$f$};
\node [bigpoint](m) at ({(2)*\InterFeuilles},{\NiveauC}) {};
\node [texto] (tm) at ({(2.3)*\InterFeuilles},{\NiveauC}) {$m$};
\node [bigpoint](p1) at ({(3.1)*\InterFeuilles},{\NiveauD}) {};
\node [texto] (Tp1) at ({(3.35)*\InterFeuilles},{\NiveauD}) {$1$};
\node [bigpoint](p2) at ({(2.4)*\InterFeuilles},{\NiveauD}) {};
\node [texto] (Tp1) at ({(2.65)*\InterFeuilles},{\NiveauD}) {$2$};
\node [bigpoint](p3) at ({(1.7)*\InterFeuilles},{\NiveauD}) {};
\node [texto] (Tp1) at ({(1.95)*\InterFeuilles},{\NiveauD}) {$3$};
\node [bigpoint](p4) at ({(1)*\InterFeuilles},{\NiveauD}) {};
\node [texto] (Tp1) at ({(1.25)*\InterFeuilles},{\NiveauD}) {$4$};

% Arcs (MODIFIABLES : Styles)

\draw[-, thick,line width = 0.4mm]  (R) to [out = -100, in = -0] (Rbis);
\draw[->, thick,line width = 0.4mm]  (Rbis) to [out = 150, in = 80] (f);
\draw[-, thick,line width = 0.4mm]  (f) to (m);
\draw[-, thick,line width = 0.4mm]  (m) to (p1);
\draw[-, thick,line width = 0.4mm]  (m) to (p2);
\draw[-, thick,line width = 0.4mm]  (m) to (p3);
\draw[-, thick,line width = 0.4mm]  (m) to (p4);

\end{tikzpicture}

\end{center}
%~\hspace{4cm} \includegraphics{Clockwise}
 \caption{Father, children, and the clockwise order}
\label{fig 2} 
\end{figure}
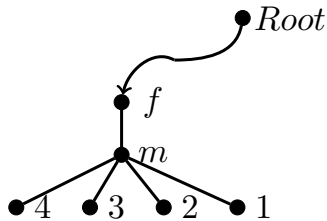

For each node $m$, there is a unique path between $m$ and the root; the depth of the node $m$ is the lenght of this path. 
The adjacent node to $m$ in this path is the father of $m$, the
other adjacent nodes are the children of $m$. In Figure~\ref{fig 2},
$f$ is the father of $m$. The children of $m$ are the nodes $1,2,3,4$;
they are ordered in the clockwise order. We will say that $1$ is
the most left (with respect to $m$) and $4$ is the most right. Of
course, when $m$ is the root, we must specify who is the first child
of $m$ in the clockwise order (or, equivalently, the most left).

\begin{figure}[h!]
\begin{center}
% Racine en Haut, développement vers le bas
\tikzset{
    ultra thin/.style= {line width=0.1pt},
    very thin/.style=  {line width=0.2pt},
    thin/.style=       {line width=0.4pt},% thin is the default
    semithick/.style=  {line width=0.6pt},
    thick/.style=      {line width=0.6pt},
    very thick/.style= {line width=1.2pt},
    ultra thick/.style={line width=1.2pt}
}
\begin{tikzpicture}[xscale=1,yscale=1]
% Styles (MODIFIABLES)
\tikzstyle{noeud} = [rectangle,rounded corners, fill = white,ultra thick,minimum height = 6.7mm, minimum width = 5mm ,text centered,font=\bfseries\fontsize{8}{0}\selectfont,draw]
\tikzstyle{invisiblepoint} = [fill = white, circle, scale = 0.01]
\tikzstyle{fleche}=[->,>=latex,ultra thick]

% Dimensions (MODIFIABLES)
\def\DistanceInterNiveaux{1.4}
\def\DistanceInterFeuilles{1.4}
% Dimensions calculées (NON MODIFIABLES)
\def\NiveauA{(-0)*\DistanceInterNiveaux}
\def\NiveauB{(-0.8)*\DistanceInterNiveaux}
\def\NiveauC{(-1.45)*\DistanceInterNiveaux}
\def\NiveauD{(-2.05)*\DistanceInterNiveaux}
\def\InterFeuilles{(1)*\DistanceInterFeuilles}
% Noeuds (MODIFIABLES : Styles et Coefficients d'InterFeuilles)
\node[noeud] (R) at ({(2)*\InterFeuilles},{\NiveauA}) {1};

\node[noeud] (B1) at ({(-0.5)*\InterFeuilles},{\NiveauB}) {13};
\node [noeud](B2) at ({(1.35)*\InterFeuilles},{\NiveauB}) {11};
\node [noeud](B3) at ({(2.65)*\InterFeuilles},{\NiveauB}) {8};
\node [noeud](B4) at ({(4.5)*\InterFeuilles},{\NiveauB}) {2};

\node [noeud](C1) at ({(-1.85)*\InterFeuilles},{\NiveauC}) {16};
\node [noeud](C2) at ({(-0.75)*\InterFeuilles},{\NiveauC}) {15};
\node [noeud](C3) at ({(0.15)*\InterFeuilles},{\NiveauC}) {14};

\node [noeud](C4) at ({(1.20)*\InterFeuilles},{\NiveauC}) {12};

\node [noeud](C5) at ({(2.30)*\InterFeuilles},{\NiveauC}) {10};
\node [noeud](C6) at ({(3)*\InterFeuilles},{\NiveauC}) {9};

\node [noeud](C7) at ({(4.25)*\InterFeuilles},{\NiveauC}) {5};
\node [noeud](C8) at ({(5.00)*\InterFeuilles},{\NiveauC}) {4};
\node [noeud](C9) at ({(5.85)*\InterFeuilles},{\NiveauC}) {3};

\node [noeud](D1) at ({(3.7)*\InterFeuilles},{\NiveauD}) {7};
\node [noeud](D2) at ({(4.65)*\InterFeuilles},{\NiveauD}) {6};

% Arcs (MODIFIABLES : Styles)

\draw[-,line width = 0.3mm]  (R.south) to (B1.north);
\draw[-,line width = 0.3mm]  (R.south) to (B2.north);
\draw[-, line width = 0.3mm]  (R.south) to (B3.north);
\draw[-, line width = 0.3mm]  (R.south) to (B4.north);

\draw[-, line width = 0.3mm]  (B1.south) to (C1.north);
\draw[-, line width = 0.3mm]  (B1.south) to (C2.north);
\draw[-, line width = 0.3mm]  (B1.south) to (C3.north);

\draw[-, line width = 0.3mm]  (B2.south) to (C4.north);

\draw[-, line width = 0.3mm]  (B3.south) to (C5.north);
\draw[-, line width = 0.3mm]  (B3.south) to (C6.north);

\draw[-, line width = 0.3mm]  (B4.south) to (C7.north);
\draw[-, line width = 0.3mm]  (B4.south) to (C8.north);
\draw[-, line width = 0.3mm]  (B4.south) to (C9.north);

\draw[-, thick,line width = 0.3mm]  (C7.south) to (D1.north);
\draw[-, thick,line width = 0.3mm]  (C7.south) to (D2.north);

\draw[fleche, line width = 0.5mm] (3.5,-0.25)--(5.8,-0.5);

\end{tikzpicture}

\end{center}
%~\hspace{1cm} \includegraphics{LDFS} 
\caption{The LDFS in the tree T}
\label{fig 3} 
\end{figure}
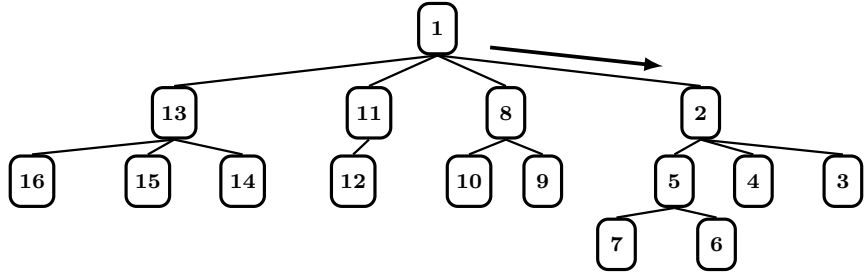

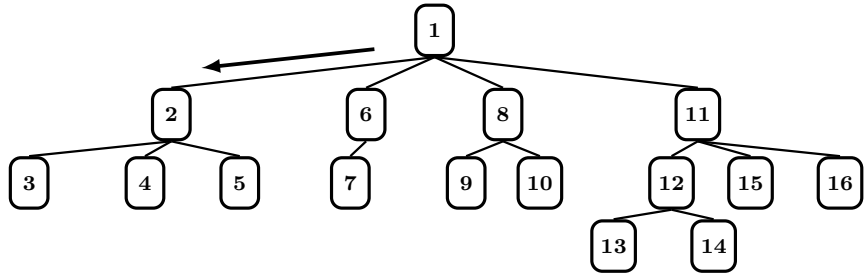
\begin{figure}[h!]
\begin{center}
% Racine en Haut, développement vers le bas
\tikzset{
    ultra thin/.style= {line width=0.1pt},
    very thin/.style=  {line width=0.2pt},
    thin/.style=       {line width=0.4pt},% thin is the default
    semithick/.style=  {line width=0.6pt},
    thick/.style=      {line width=0.6pt},
    very thick/.style= {line width=1.2pt},
    ultra thick/.style={line width=1.2pt}
}
\begin{tikzpicture}[xscale=1,yscale=1]
% Styles (MODIFIABLES)
\tikzstyle{noeud} = [rectangle,rounded corners, fill = white,ultra thick,minimum height = 6.7mm, minimum width = 5mm ,text centered,font=\bfseries\fontsize{8}{0}\selectfont,draw]
\tikzstyle{invisiblepoint} = [fill = white, circle, scale = 0.01]
\tikzstyle{fleche}=[->,>=latex,ultra thick]

% Dimensions (MODIFIABLES)
\def\DistanceInterNiveaux{1.4}
\def\DistanceInterFeuilles{1.4}
% Dimensions calculées (NON MODIFIABLES)
\def\NiveauA{(-0)*\DistanceInterNiveaux}
\def\NiveauB{(-0.8)*\DistanceInterNiveaux}
\def\NiveauC{(-1.45)*\DistanceInterNiveaux}
\def\NiveauD{(-2.05)*\DistanceInterNiveaux}
\def\InterFeuilles{(1)*\DistanceInterFeuilles}
% Noeuds (MODIFIABLES : Styles et Coefficients d'InterFeuilles)
\node[noeud] (R) at ({(2)*\InterFeuilles},{\NiveauA}) {1};

\node[noeud] (B1) at ({(-0.5)*\InterFeuilles},{\NiveauB}) {2};
\node [noeud](B2) at ({(1.35)*\InterFeuilles},{\NiveauB}) {6};
\node [noeud](B3) at ({(2.65)*\InterFeuilles},{\NiveauB}) {8};
\node [noeud](B4) at ({(4.5)*\InterFeuilles},{\NiveauB}) {11};

\node [noeud](C1) at ({(-1.85)*\InterFeuilles},{\NiveauC}) {3};
\node [noeud](C2) at ({(-0.75)*\InterFeuilles},{\NiveauC}) {4};
\node [noeud](C3) at ({(0.15)*\InterFeuilles},{\NiveauC}) {5};

\node [noeud](C4) at ({(1.20)*\InterFeuilles},{\NiveauC}) {7};

\node [noeud](C5) at ({(2.30)*\InterFeuilles},{\NiveauC}) {9};
\node [noeud](C6) at ({(3)*\InterFeuilles},{\NiveauC}) {10};

\node [noeud](C7) at ({(4.25)*\InterFeuilles},{\NiveauC}) {12};
\node [noeud](C8) at ({(5.00)*\InterFeuilles},{\NiveauC}) {15};
\node [noeud](C9) at ({(5.85)*\InterFeuilles},{\NiveauC}) {16};

\node [noeud](D1) at ({(3.7)*\InterFeuilles},{\NiveauD}) {13};
\node [noeud](D2) at ({(4.65)*\InterFeuilles},{\NiveauD}) {14};

% Arcs (MODIFIABLES : Styles)

\draw[-,line width = 0.3mm]  (R.south) to (B1.north);
\draw[-,line width = 0.3mm]  (R.south) to (B2.north);
\draw[-, line width = 0.3mm]  (R.south) to (B3.north);
\draw[-, line width = 0.3mm]  (R.south) to (B4.north);

\draw[-, line width = 0.3mm]  (B1.south) to (C1.north);
\draw[-, line width = 0.3mm]  (B1.south) to (C2.north);
\draw[-, line width = 0.3mm]  (B1.south) to (C3.north);

\draw[-, line width = 0.3mm]  (B2.south) to (C4.north);

\draw[-, line width = 0.3mm]  (B3.south) to (C5.north);
\draw[-, line width = 0.3mm]  (B3.south) to (C6.north);

\draw[-, line width = 0.3mm]  (B4.south) to (C7.north);
\draw[-, line width = 0.3mm]  (B4.south) to (C8.north);
\draw[-, line width = 0.3mm]  (B4.south) to (C9.north);

\draw[-, thick,line width = 0.3mm]  (C7.south) to (D1.north);
\draw[-, thick,line width = 0.3mm]  (C7.south) to (D2.north);

\draw[fleche, line width = 0.5mm] (2,-0.25)--(-0.3,-0.5);

\end{tikzpicture}

\end{center}
%~\hspace{-2cm} \includegraphics{RDFS} 
\caption{The RDFS in the tree T}
\label{fig 4} 
\end{figure}

The way we will visit the tree is the Depth First Search (DFS). 
There are two DFS, the Left DFS (LDFS) and the Right DFS
(RDFS). They start from the root. From a node $m$, the LDFS goes to
the most left non visited child of $m$ if it exists and otherwise 
returns to the father of $m$, while the RDFS goes to the most right
non visited child of $m$ if it exists. Figures~\ref{fig 3} and~\ref{fig 4}
give the two DFS for the tree T  in Figure~\ref{fig 1}.

\subsubsection{Prefix Tree}
\begin{figure}[h!]
\begin{center}
% Racine en Haut, développement vers le bas
\tikzset{
    ultra thin/.style= {line width=0.1pt},
    very thin/.style=  {line width=0.2pt},
    thin/.style=       {line width=0.4pt},% thin is the default
    semithick/.style=  {line width=0.6pt},
    thick/.style=      {line width=0.6pt},
    very thick/.style= {line width=1.2pt},
    ultra thick/.style={line width=1.1pt}
}
\begin{tikzpicture}[xscale=1,yscale=1]
% Styles (MODIFIABLES)
\tikzstyle{noeud} = [rectangle,fill = white,ultra thick,minimum height = 5mm, minimum width = 8mm ,text centered,font=\bfseries\fontsize{8}{0}\selectfont,draw]

% Dimensions (MODIFIABLES)
\def\DistanceInterNiveaux{1.4}
\def\DistanceInterFeuilles{1.4}
% Dimensions calculées (NON MODIFIABLES)
\def\NiveauA{(-0)*\DistanceInterNiveaux}
\def\NiveauB{(-0.6)*\DistanceInterNiveaux}
\def\NiveauC{(-1.2)*\DistanceInterNiveaux}
\def\NiveauD{(-1.8)*\DistanceInterNiveaux}
\def\NiveauE{(-2.4)*\DistanceInterNiveaux}
\def\InterFeuilles{(1)*\DistanceInterFeuilles}
% Noeuds (MODIFIABLES : Styles et Coefficients d'InterFeuilles)
\node[noeud] (R) at ({(2.5)*\InterFeuilles},{\NiveauA}) {($\emptyset$)};

\node[noeud] (B1) at ({(-0.5)*\InterFeuilles},{\NiveauB}) {(1)};
\node [noeud](B2) at ({(1.45)*\InterFeuilles},{\NiveauB}) {(2)};
\node [noeud](B3) at ({(3.6)*\InterFeuilles},{\NiveauB}) {(3)};
\node [noeud](B4) at ({(4.7)*\InterFeuilles},{\NiveauB}) {(4)};

\node [noeud](C1) at ({(-1.85)*\InterFeuilles},{\NiveauC}) {(1,2)};
\node [noeud](C2) at ({(-0.75)*\InterFeuilles},{\NiveauC}) {(1,3)};
\node [noeud](C3) at ({(0.15)*\InterFeuilles},{\NiveauC}) {(1,4)};

\node [noeud](C4) at ({(1.20)*\InterFeuilles},{\NiveauC}) {(2,3)};
\node [noeud](C5) at ({(2.30)*\InterFeuilles},{\NiveauC}) {(2,4)};

\node [noeud](C6) at ({(3.8)*\InterFeuilles},{\NiveauC}) {(3,4)};

\node [noeud](D1) at ({(-2.7)*\InterFeuilles},{\NiveauD}) {(1,2,3)};
\node [noeud](D2) at ({(-1.6)*\InterFeuilles},{\NiveauD}) {(1,2,4)};
\node [noeud](D3) at ({(-0.6)*\InterFeuilles},{\NiveauD}) {(1,3,4)};

\node [noeud](D4) at ({(1)*\InterFeuilles},{\NiveauD}) {(2,3,4)};

\node [noeud](E1) at ({(-2.7)*\InterFeuilles},{\NiveauE}) {(1,2,3,4)};

% Arcs (MODIFIABLES : Styles)

\draw[-,line width = 0.35mm]  (R.south) to (B1.north);
\draw[-,line width = 0.35mm]  (R.south) to (B2.north);
\draw[-, line width = 0.35mm]  (R.south) to (B3.north);
\draw[-, line width = 0.35mm]  (R.south) to (B4.north);

\draw[-, line width = 0.35mm]  (B1.south) to (C1.north);
\draw[-, line width = 0.35mm]  (B1.south) to (C2.north);
\draw[-, line width = 0.35mm]  (B1.south) to (C3.north);

\draw[-, line width = 0.35mm]  (B2.south) to (C4.north);
\draw[-, line width = 0.35mm]  (B2.south) to (C5.north);

\draw[-, line width = 0.35mm]  (B3.south) to (C6.north);

\draw[-, line width = 0.35mm]  (C1.south) to (D1.north);
\draw[-, line width = 0.35mm]  (C1.south) to (D2.north);

\draw[-, line width = 0.35mm]  (C2.south) to (D3.north);

\draw[-, line width = 0.35mm]  (C4.south) to (D4.north);

\draw[-, line width = 0.35mm]  (D1.south) to (E1.north);

\end{tikzpicture}

\end{center}
\caption{Prefix tree with 4 items}
\label{fig 5} 
\end{figure}
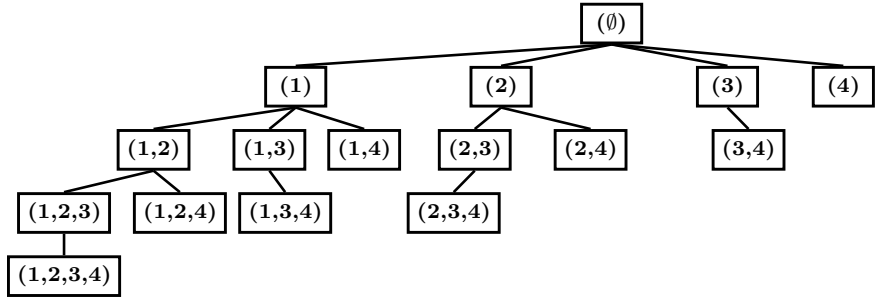

The tree enumeration of the itemsets is based on the order of the
items. Let $I=\{1,2,\ldots,n\}$ be the ordered set of the items; a $k$-itemset
is represented by a vector with $k$ increasing integers coordinates
in $\{1,2,\ldots,n\}$. We define the prefix tree, $pT_{n}$, a tree
whose nodes are the itemsets, as follows. We proceed by defining successively each generation. This ensures that we get a planar tree. 
\begin{itemize}
 \item The root of $pT_{n}$ is the empty itemset $(\emptyset)$.
\item The children of $(\emptyset)$ are the
1-itemsets $(1),(2),\ldots,(n)$ respecting their order and such that
$(1)$ is the most right (with respect to the root) and $(n)$ is
the most left. 
\item The children of the node $A=(i_{1},\ldots,i_{k})$ are the nodes $(i_{1},\ldots,i_{k},j) $ where $i_{k}<j\leq n$ and  
 $j$ is ordered from the right (of $A$) to the left. More precisely, $(i_{1},\ldots,i_{k},i_{k}+1)$ is the most right child of $A$ and  $(i_{1},\ldots,i_{k},n)$ is the most left. Of course if $i_{k}=n$, then $A$ is a leaf of the tree.
\end{itemize}
Figure~\ref{fig 5} gives $pT_{4}$.

\begin{prop} \label{prop 1} For a node $a=(i_{1},\ldots,i_{k})$
in $pT_{n}$, $h(a)=k$ is the depth  of $a$. \end{prop}

\paragraph{Proof}
 The nodes $(1),(2),\ldots,(n)$ are
the children of the root $(\emptyset)$ and thus their depth is equal
to $1$. Assume the proposition true for $k-1\geq1$, and let $a=(i_{1},\ldots,i_{k})$;
$a$ is a child of the node $r=(i_{1},\ldots,i_{k-1})$ and thus $h(a)=h(r)+1$.
Since $h(r)=k-1$ by assumption,  the result follows.

\medskip{}

From the definition of the prefix tree, it is easy to deduce the following proposition.

\begin{prop} \label{leaf} Let $pT_{n}$ be the prefix tree for the set of items $I=\{1,2,\ldots,n\}.$ The following assertions are equivalent:

(i) $a=(i_{1},\ldots,i_{k})$ is a leaf, 

(ii) $i_{k}=n$,  

(iii) $a=(i_{1},\ldots,i_{k})$ is the most left child of $(i_{1},\ldots,i_{k-1})$.

\end{prop}

We have the following obvious property.

\begin{property} \label{property 1} Let $m=(i_{1},\ldots,i_{k})$ be an itemset and 
$pT_{n}(m)$ be the subtree rooted in the node $m$ in $pT_{n}$. The set of the nodes of $pT_{n}(m)$ is composed of $m$ and of all the nodes $a=(j_{1},\ldots,j_{l})$ such that $l>k$ and $(j_{1},\ldots,j_{k})=(i_{1},\ldots,i_{k})$. 
In particular, $m\subset a $ for each node $a$ of $pT_{n}(m)$.
\end{property}

\subsubsection{Frequent prefix tree}

 Let us first note that we have the following important property:
\begin{equation}
\label{apriori}
  \mbox{if} \  A \subset B \ \mbox{then} \  \{ \freq(A)< \sigma   \Rightarrow  \freq(B)< \sigma  \} .
  \end{equation}
In other words, if $B$ is a frequent itemset, then all the itemsets included in $B$ are frequent.   
 
Since $\freq(\emptyset)=1$, $(\emptyset)$ is always frequent. In view of (\ref{apriori})  and  of Property \ref{property 1}, if
a node $m$ of $pT_{n}$  is not frequent, then none of the nodes in the subtree rooted in
the node $m$ is frequent. Consequently, by removing all the nodes $a$ such that $\freq(a)<\sigma$ in $pT_{n}$, we obtain a tree with $(\emptyset)$ as root.
We can thus  define  the frequent prefix tree as follows. 

\begin{defi} Let $\sigma$ be the minSup. We call frequent prefix tree, the tree $pT_{n}^{\sigma}$ obtained
from $pT_{n}$ by removing the nodes $a$ such that $\freq(a)<\sigma$.
\end{defi}

\begin{prop} \label{prop 2} If $pT_{n}^{\sigma}(m)$ is the subtree  rooted in  the node $m$ in the frequent prefix tree $pT_{n}^{\sigma}$, then $m$ is frequent and $pT_{n}^{\sigma}(m)$  is obtained by  removing from $pT_{n}(m)$ all the nodes $a$ such that $\freq(a)<\sigma$.
\end{prop}

\begin{rque} \label{Remark1bis}
If $m$ is not frequent, then $m$ is not in $pT_{n}^{\sigma}$ and  $pT_{n}^{\sigma}(m)$ is empty.
\end{rque}

\begin{rque} \label{Remarkter}
 If $m$ is not in  $pT_{n}^{\sigma}$, then none of the nodes of $pT_{n}(m)$ are in $pT_{n}^{\sigma}$. 
\end{rque}

\begin{rque} \label{Remark1}
By Property \ref{property 1}, we have $m\subset a $ for each node $a$ of $pT_{n}^{\sigma}(m).$ 
\end{rque}

\subsection{Recursive Building of the Prefix Trees}

 \subsubsection{The Recurrence relation between the prefix trees}
 
Let $a=(i_{1},\ldots,i_{k})$ be an itemset, let $j$ be an item such
that $j>i_{k}$; we set $(a,j)=(i_{1},\ldots,i_{k},j)$. In the particular
case $a=(\emptyset)$, we set $((\emptyset),j)=(j)$.

Let $pT_{0}$ be the tree with one node ($\emptyset)$. We recall that $pT_{k}$ is the prefix tree based on the items $\{1,\ldots,k\}$. As an example, $pT_{1}$ is a tree
 with two nodes, $(\emptyset)$ and $(1)$, where $(\emptyset)$ is the root and $(1)$ is its child.

\begin{prop} \label{prop 3}

The leaves of $pT_{k+1}$ are the nodes $(a, k+1)$ where $a$ is a node of $pT_{k}$. The tree $pT_{k}$ is obtained from $pT_{k+1}$ by removing all the leaves $(a, k+1)$.  

\end{prop}

\paragraph{Proof}
From Proposition  \ref{leaf}, the set of leaves of $pT_{k+1}$ is exactly the set of nodes $(a, k+1)$ where $a$ is an itemset based on ${1,\ldots,k}$, that is, a node of $pT_{k}$. The edge relation between such itemsets $a$ is the same in $pT_{k}$ and in $pT_{k+1}$. The result follows. \\

\begin{figure}[h!]
\begin{center}
% Racine en Haut, développement vers le bas

\begin{tikzpicture}[xscale=1,yscale=1]
% Styles (MODIFIABLES)

\tikzstyle{bigpoint} = [fill = black, circle, draw, scale = 0.5]
\tikzstyle{texto} = [thick, font=\bfseries\fontsize{8}{0}\selectfont]
\tikzstyle{invisiblepoint} = [fill = black, circle, scale = 0.01, draw]

% Dimensions (MODIFIABLES)
\def\DistanceInterNiveaux{1.5}
\def\DistanceInterFeuilles{1.5}
% Dimensions calculées (NON MODIFIABLES)
\def\NiveauA{(-0)*\DistanceInterNiveaux}
\def\NiveauB{-0.4)*\DistanceInterNiveaux}
\def\NiveauC{(-0.8)*\DistanceInterNiveaux}
\def\NiveauD{(-1.8)*\DistanceInterNiveaux}
\def\NiveauE{(-2.2)*\DistanceInterNiveaux}
\def\NiveauF{(-2.6)*\DistanceInterNiveaux}
\def\NiveauG{(-3.0)*\DistanceInterNiveaux}
\def\InterFeuilles{(1)*\DistanceInterFeuilles}
% Noeuds (MODIFIABLES : Styles et Coefficients d'InterFeuilles)
\node[bigpoint] (F11) at ({(0)*\InterFeuilles},{\NiveauA}) {};
\node [texto] (F11t) at ({(0.3)*\InterFeuilles},{\NiveauA}) {($\emptyset$)};
\node[bigpoint] (F12) at ({(0)*\InterFeuilles},{\NiveauB}) {};
\node [texto] (F12t) at ({(0.3)*\InterFeuilles},{\NiveauB}) {(1)};

\node[bigpoint] (F21) at ({(5)*\InterFeuilles},{\NiveauA}) {};
\node [texto] (F21t) at ({(5.3)*\InterFeuilles},{\NiveauA}) {($\emptyset$)};
\node[bigpoint] (F22) at ({(4.7)*\InterFeuilles},{\NiveauB}) {};
\node [texto] (F22t) at ({(5)*\InterFeuilles},{\NiveauB}) {(1)};
\node[bigpoint] (F23) at ({(5.8)*\InterFeuilles},{\NiveauB}) {};
\node [texto] (F23t) at ({(6.1)*\InterFeuilles},{\NiveauB}) {(2)};
\node[bigpoint] (F24) at ({(4.5)*\InterFeuilles},{\NiveauC}) {};
\node [texto] (F24t) at ({(4.8)*\InterFeuilles},{\NiveauC}) {(1,2)};

\node[bigpoint] (F31) at ({(1.5)*\InterFeuilles},{\NiveauD}) {};
\node [texto] (F31t) at ({(1.9)*\InterFeuilles},{\NiveauD}) {($\emptyset$)};
\node[bigpoint] (F32) at ({(1)*\InterFeuilles},{\NiveauE}) {};
\node [texto] (F32t) at ({(1.3)*\InterFeuilles},{\NiveauE}) {(1)};
\node[bigpoint] (F33) at ({(2.5)*\InterFeuilles},{\NiveauE}) {};
\node [texto] (F33t) at ({(2.8)*\InterFeuilles},{\NiveauE}) {(2)};
\node[bigpoint] (F34) at ({(3.5)*\InterFeuilles},{\NiveauE}) {};
\node [texto] (F34t) at ({(3.8)*\InterFeuilles},{\NiveauE}) {(3)};

\node[bigpoint] (F35) at ({(0)*\InterFeuilles},{\NiveauF}) {};
\node [texto] (F35t) at ({(0.4)*\InterFeuilles},{\NiveauF}) {(1,2)};
\node[bigpoint] (F36) at ({(1.3)*\InterFeuilles},{\NiveauF}) {};
\node [texto] (F36t) at ({(1.7)*\InterFeuilles},{\NiveauF}) {(1,3)};
\node[bigpoint] (F37) at ({(2.7)*\InterFeuilles},{\NiveauF}) {};
\node [texto] (F37t) at ({(3.1)*\InterFeuilles},{\NiveauF}) {(2,3)};

\node[bigpoint] (F38) at ({(0)*\InterFeuilles},{\NiveauG}) {};
\node [texto] (F38t) at ({(0.5)*\InterFeuilles},{\NiveauG}) {(1,2,3)};

\node [texto] (Txt1) at ({(0.5)*\InterFeuilles},{(-0.8)*\DistanceInterNiveaux}) {$pT_1=\emptyset+(1)$};

\node [texto] (Txt1b) at ({(0.85)*\InterFeuilles},{(-1.1)*\DistanceInterNiveaux}) {$\mbox{LDFS}(1)=((\emptyset),(1))$};

\node [texto] (Txt2) at ({(5.3)*\InterFeuilles},{(-1.2)*\DistanceInterNiveaux}) {$pT_2 = pT_1 + (2)$ };

\node [texto] (Txt3) at ({(5.80)*\InterFeuilles},{(-1.5)*\DistanceInterNiveaux}) {$\mbox{LDFS}(2)=((\emptyset),(2),(1),(1,2))$};

\node [texto] (Txt4) at ({(0.8)*\InterFeuilles},{(-3.4)*\DistanceInterNiveaux}) {$pT_3 = pT_2 + (3) $};
\node [texto] (Txt2) at ({(2.4)*\InterFeuilles},{(-3.7)*\DistanceInterNiveaux}) {$\mbox{LDFS}(3)=((\emptyset),(3),(2),(2,3),(1),(1,3), (1,2),(1,2,3))$};

% Arcs (MODIFIABLES : Styles)

\draw[-,line width = 0.45mm]  (F11) to (F12);

\draw[-,line width = 0.4mm]  (F21) to (F22);
\draw[-,line width = 0.4mm]  (F21) to (F23);
\draw[-,line width = 0.4mm]  (F22) to (F24);

\draw[-,line width = 0.4mm]  (F31) to (F32);
\draw[-,line width = 0.4mm]  (F31) to (F33);
\draw[-,line width = 0.4mm]  (F31) to (F34);

\draw[-,line width = 0.4mm]  (F32) to (F35);
\draw[-,line width = 0.4mm]  (F32) to (F36);
\draw[-,line width = 0.4mm]  (F33) to (F37);
\draw[-,line width = 0.45mm]  (F35) to (F38);

\end{tikzpicture}

\end{center}
%~\hspace{2cm} \includegraphics{recursivePtree} 
\caption{Recursive prefix tree $k=1,2,3$ and their LDFS}
\label{fig 6} 
\end{figure}
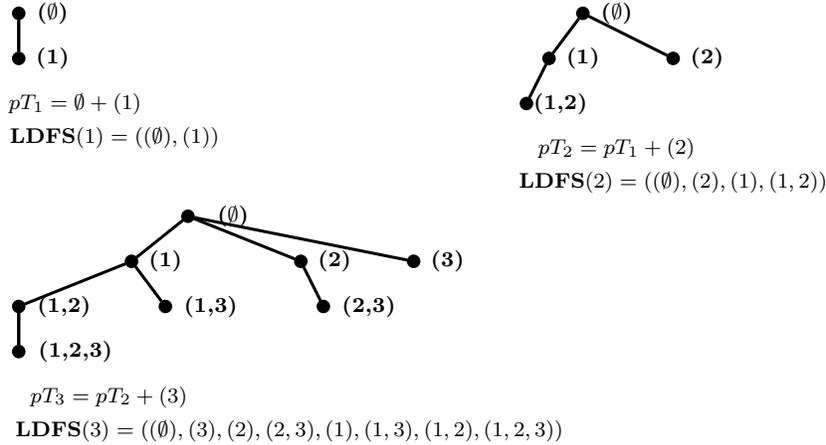

As a consequence, we have the following corollary, which gives the way to build the tree  $pT_{n}$  recursively (as an illustration, see Figure~\ref{fig 6}). 

\begin{cor} \label{build}  
The tree $pT_{k+1}$ is built from $pT_{k}$  by adding $(a,k+1)$ as the most left child of $a$ for each node $a$ in $pT_{k}$. 

\end{cor}

We can now state the following proposition, which  gives the way to build the frequent tree  $pT_{n}^{\sigma}$  recursively. 

\begin{prop} [Recursion] \label{buildfreq}
Let $\sigma$ be the minSup. The tree $pT_{k+1}^{\sigma}$ is built from $pT_{k}^{\sigma}$ as follows. For each node $a$ in $pT_{k}^{\sigma}$, if $(a,k+1)$ is frequent, then it is added  as the most left child of $a$.

\end{prop}

\paragraph{Proof} 
$pT_{k+1}$ is obtained from $pT_{k}$ by adding the nodes $(a,k+1)$ with $a$ in $pT_{k}$. 
$pT_{k+1}^{\sigma}$ is obtained  from $pT_{k+1}$ by removing all the non frequent nodes. 
Equivalently, to obtain $pT_{k+1}^{\sigma}$  we can first remove all the non frequent nodes of $pT_{k}$ and then add only the frequent nodes $(a,k+1)$. 

\subsubsection{Ordering the nodes of  $pT_{k}$ and $pT_{k}^{\sigma}$ }

For the construction of $pT_{k+1}^{\sigma}$ we need to visit the nodes $a$ of $pT_{k}^{\sigma}$ and to test if $(a,k+1)$ is frequent. In fact, due to the following proposition, we do not need to visit all the nodes $a$. 
  
\begin{prop} [Pruning] \label{Pruning}
Let $\sigma$ be the minSup and let $m$ be a node of $pT_{k}^{\sigma}$. If $\freq((m,k+1))< \sigma$, then  $\freq((b,k+1))< \sigma$ for all the nodes $b$ in the subtree $pT_{k}^{\sigma}(m)$. 

\end{prop}

\paragraph{Proof} 
In view of Remark~\ref{Remark1}, $m\subset b$. Thus $(m,k+1) \subset (b,k+1)$  and $\freq((b,k+1)) \leq \freq((m,k+1)).$ 

\begin{rque} \label{Remark synthetic}
Let $m$ be a node in $pT_{k}$.
 If $m$ is in $pT_{k}^{\sigma}$ and $(m,k+1)$ is not in $pT_{k+1}^{\sigma}$, then, for any node $a$ in $pT_{k}(m)$, $(a,k+1)$ 
is not in $pT_{k+1}^{\sigma}$. 
\end{rque}

In view of Proposition \Ref{Pruning}, we must visit the nodes of $pT_{k}^{\sigma}$ in such a way that each node $m$ is visited before all the other nodes of its subtree $pT_{k}^{\sigma}(m)$. Among the most well known tree-traversing that satisfy the above, there are the Breadth First Search (BDS) and the Depth First Search (DFS). For reasons of efficiency in the implementation of our algorithm, we choose to use the Depth First Search traversal. Moreover, to keep our algorithm recursive, we need to simultaneously carry out the search of the frequent itemsets and the search of the association rules. We are then led to use the Left Depth First Search traversal (LDFS).

\begin{prop} \label{Proposition 3} Let  $r$ be a node in $pT_{k}$ (respectively, in $pT_{k}^{\sigma}$). 
If a node $m$ is such that $m\subset r$, then $m$ is before $r$ in $pT_{k}$ (respectively, in $pT_{k}^{\sigma}$), with respect to the LDFS order. In particular $m$ is in $pT_{k}$ (respectively, in $pT_{k}^{\sigma}$).
 \end{prop}

\paragraph{Proof}

We first prove Proposition~\ref{Proposition 3}  by induction in the case $r$ is in $pT_{k}$. The
property is obvious for $pT_{1}$. From the construction of $pT_{k+1}$,
we have the following two properties: 
\begin{itemize}
\item[$(i)$] for any $m$ in $pT_{k}$, $(m,k+1)$ comes after $m$ in $pT_{k+1}$. 
\item[$(ii)$] if $m$ is before $r$ in $pT_{k}$, then $(m,k+1)$ is before $(r,k+1)$
in $pT_{k+1}$. 
\end{itemize}
Asssume Proposition~\ref{Proposition 3} is true for  $pT_{k}$. Let $r$
in $pT_{k+1}$ and not in $pT_{k}$. Then $r=(v,k+1)$ with $v$
in $pT_{k}$. Let a node $m$ be such that $m\subset r$. Either
$m$ is a subset of $v$ and thus it is before $v$ (by induction
assumption), which is before $r$ by property $(i)$. Or $m=(w,k+1)$
where $w$ is a subset of $v$ and thus $w$ is before $v$ in $pT_{k}$
(by induction assumption); it follows from $(ii)$ that $m$ is before
$r$ in $pT_{k+1}$.

Let us now consider the case $r$ is in $pT_{k}^{\sigma}$. Then, $r$ is in  $pT_{k}$ and thus $m$ is before $r$ in  $pT_{k}$. Since $r$
is frequent and since $m\subset r$, $m$ also is  frequent  and thus $m$
is before $r$ in  $pT_{k}^{\sigma}$.

\begin{rque} \label{Remark 2}
 Proposition~\ref{Proposition 3} is the key to ensure that we can simultaneously carry out the mining of the frequent itemsets and the association rules. Indeed, it says that the order is such that, when a frequent itemset $C$ is found, all the $A\subset C$ have been tested before $C$ and their relative support have been computed. 
\end{rque}

\begin{rque} \label{Remark 3} Clearly Proposition~\ref{Proposition 3} does not hold when the nodes   in $pT_{k}$  
are ordered with respect to the RDFS order. For example, the node $(1,2,3)$ is then before its subsets $((1,3),(2),(2,3),(3))$
in $pT_{3}$ . 
\end{rque}
Before starting the presentation of the algorithm we need the following definition.

\begin{defi} \label{mark}
Let $a=(j_{1},\ldots, j_{l-1}, j_{l})$ be a node. If $j_{l}>j_{l-1}+1$, we set $H(a)=\{b=(j_{1},\ldots, j_{l-1}, s ,j_{l}),  j_{l-1}<s< j_{l}\} $. If $j_{l}=j_{l-1}+1$, $H(a)=\emptyset $.
\end{defi}

We note that if $a$ is not frequent, all the nodes in $H(a)$ are not frequent. Note also that the nodes in $H(a)$ are children of brothers of $a$. 

\section{{The Recursive Algorithm, its Updating, and its Applications}}
\label{section algo}

{We first present both versions of our algorithm: the first one, the PrefRec FIM algorithm, only gives the frequent
itemsets, whereas the second one, the PrefRec ARM algorithm, also gives the association rules. Then, we explain how the algorithm PrefRec can be updated, and finally give its applications.}

\subsection{The PrefRec FIM Algorithm} 

The nodes $a$ of a tree $T$ are represented by a vector with increasing
integer coordinates $a=(j_{1},\ldots,j_{l})$. To $a$ is associated
the observation vector 
\[
(z_{i}^{(a)})_{1\leq i\leq N}=(x_{i}^{(j_{1})}\times\ldots\times x_{i}^{(j_{l})})_{1\leq i\leq N}\in\{0,1\}^{N}
\]
and its support $\supp(a)=\sum_{1\leq i\leq N}z_{i}^{(a)}$.
The root of $T$ is always the empty vector $(\emptyset)$ ; it is
associated to $1_{N}=(1,1,\ldots ,1)\in\{0,1\}^{N}$ and $\supp((\emptyset))=N$.
The tree $T$ is visited with the help of the LDFS; this gives a total
order on the set of nodes.  
The PrefRec FIM algorithm is based  on Proposition~\ref{buildfreq}
and Proposition~\ref{Pruning}. For each $k$, $1\leq k\leq n$,
this algorithm gives the tree $pT_{k}^{\sigma}$ whose nodes are the frequent
itemsets based on the items $\{1,\ldots,k\}$. The algorithm consists of successive applications of the FIM-Traversal function.

\paragraph{\textit{PrefRec FIM Algorithm}} 
\begin{itemize}
\item[1-] $M=\{(k),1\leq k\leq n\}$ is the set of given 1-items and each
 $(k)$ is associated to the observation vector $(x_{i}^{(k)})_{1\leq i\leq N}$
\item[2-] $\sigma$ is the minSup
\item[3-] $T_{0}=tree\ with\ one\ node=\{(\emptyset)\}$
\item[4-] \textbf{For} $k=1\ to\ n$  \textbf{do}\\  
 $T_{k}=\FIMTrav(T_{k-1},(k),\sigma)$
 
\end{itemize}

\paragraph{\textit{FIM-Traversal function}}

The $\FIMTrav$ function takes as arguments: 
\begin{itemize}
\item $\sigma$ the minSup 
\item $(k)\in M$ 
\item A tree $T$ with root $(\emptyset)$ and whose nodes are itemsets  $a=(j_{1},\ldots,j_{l})$, $j_l<k$, associated to the observation vectors $(z_{i}^{(a)})_{1\leq i\leq N}$. The support of each node $a$
is given and satisfies $\supp(a)\geq\sigma N$.
\item To each node $a$ of  $T$ is associated a value $v(a)\in \{0,1\}$.
\end{itemize}
The function returns a tree $T'$ with its LDFS and whose set
of nodes contains the set of nodes of $T$ with order preserving.
The function $\FIMTrav(T,(k),\sigma)$ works as follows: 
\begin{itemize}
\item[1-] Initialize $v(a)=1$ for all nodes $a$ of $T$
\item[2-] Traverse the nodes of $T$ in the LDFS order
\item[3-] \textbf{while} the current node $a$ has not been visited before, \textbf{do} \\
\textbf{if} $v(a)=1$ \textbf{then} \\
 compute 
$\supp((a,k))=\sum_{1\leq i\leq N} (z_{i}^{(a)} \times  x_{i}^{(k)})$ \\
 \textbf{if} $\supp((a,k))\geq\sigma N$  \textbf{then} \\
 add to $T$ the node $(a,k)$ as the
most left child of $a$, save $\supp((a,k))$, and go to the next node after $a$ \\
 \textbf{else} \\
 For all $m$ in $H(a)$ set $v(m)=0$ and go to the first node $b$ such that
$b\notin T(a)$ where $T(a)$ is the subtree of $T$ rooted in the node $a$ \\
\textbf{end if} \\
 \textbf{else} \\
go to the first node $b$ such that
$b\notin T(a)$ where $T(a)$ is the subtree of $T$ rooted in the node $a$ \\
 \textbf{end if} \\
\textbf{end while} 

\item[4-] At the end of this function, the obtained tree is $\FIMTrav(T,(k),\sigma)$. 
\end{itemize}

The heart of the algorithm is the $\FIMTrav$ function; it is the key of the recursiveness. When the set 
$M$ of items is not given, but the items come successively, the algorithm
PrefRec starts with an initial tree (e.g.
$(\emptyset)$) of frequent itemsets and then applies the $\FIMTrav$
function to each new item $(k)$.  

\subsection{The PrefRec ARM Algorithm}

\subsubsection{The RULE function}

We first define the function $\Rule$. This function
takes as arguments a frequent itemset $C$ and a minConf $\tau$ and
returns the set of association rules $\Rule(C,\tau)=\{R(A,C),A\subset C,\conf(R(A,C))\geq\tau$\}.
More precisely, the function returns a tree $T_C$  whose nodes are the itemsets $A\subset C$  such that $\conf(R(A,C))=\supp(C)/\supp(A)\geq\tau$. The building of $T_C$ uses the following pruning property 
$$B \subset A \subset C \mbox{ and }\conf(R(A,C))<\tau \Rightarrow \conf(R(B,C))<\tau.$$ 
Of course the function gives also the values $\conf(R(A,C))$ for each node A of $T_C$. We use the RDFS but we may also use the LDFS.

\paragraph{\textit{The RULE function}}
\begin{itemize}
\item[1-] $T_C=tree\ with\ one\ node=\{C\}$ and $C=(i_{1},\ldots ,i_{m})$
\item[2-] \textit{Step 1} For each $A\subset C$ with size $m-1$, if $\supp(C)/\supp(A)\geq\tau$,
add $A$ in $T_C$ as a child of $C$.
\item[3-] \textit{Step 2} Traverse the nodes of $T_C$ in the RDFS order \\
\textbf{while} the current node $a$ has not been visited before, \textbf{do} \\
\textbf{if} $a$ is 1-itemset \textbf{then} \\
 go to the next node after $a$ in the RDFS of $T_C$ \\
\textbf{else} \\
add in $T_C$ as child of $a$  all the $u=a\cap b$ where $b$ is brother of $a$ in the left of $a$ satisfying $\supp(C)/\supp(a\cap b)\geq\tau$ and  go to the next node after $a$ in the RDFS of $T_C$ \\
\textbf{end if} \\
\textbf{end while}

\item[4-] At the end of the $\Rule$ function, the obtained tree is $T_C=\Rule(C,\tau)$. 
\end{itemize}

\subsubsection{The PrefRec ARM Algorithm}

For each  $k$, $1\leq k\leq n$, the PrefRec ARM algorithm  gives the tree $pT_{k}^{\sigma}$ whose nodes are the frequent
itemsets based on the items $\{1,\ldots,k\}$, as well as  the set $\Rule(C, \tau)$  for each node $C$ in $pT_{k}^{\sigma}$.

\paragraph{\textit{PrefRec ARM Algorithm}} ~ \\
In PrefRec FIM algorithm, replace the  $\FIMTrav$ function by the 
$\ARMTrav$ function.

\paragraph{\textit{The ARM-Traversal function}}
The $\ARMTrav$ function takes as arguments those of  the  $\FIMTrav$ function increased by the minConf $\tau$.\\ 
The function returns a  tree T' with its LDFS and whose set
of nodes contains the set of nodes of T with order preserving and
gives, for each new node A, the set $\Rule(A,\tau)$.\\ 
The function $\ARMTrav(T,(k),\sigma,\tau)$
works as the  $\FIMTrav$ function where the statements
\begin{itemize}
\item
 \textbf{if} $\supp((a,k))\geq\sigma N$  \textbf{then} \\
 add to $T$ the node $(a,k)$ as the
most left child of $a$, save $\supp((a,k))$, and go to the next node after $a$ 
\end{itemize}
are replaced by 
\begin{itemize}
\item
\textbf{if} $\supp(a,k)\geq\sigma N$ \textbf{then} \\
 add to $T$ the node $(a,k)$ as the
most left child of $a$,  save $\supp(a,k)$, compute $\Rule((a,k),\tau)$, and go to the next node after $a$ 
\end{itemize}

Of course, when the database $M=\{(k),1\leq k\leq n\}$ is fixed, instead of applying PrefRec ARM Algorithm, we can first apply PrefRec FIM Algorithm and then apply the Rule function to all the frequent itemsets found.

\subsection{Updating}

At each step $n$, we can introduce a new item and/or 
remove the first item; we consider both operations in the FIM case,
the ARM case being easily deduced. The  minSup value is still
$\sigma$.

\subsubsection{Adding a new item}

After doing  step  $(n-1)$, we got
the tree of frequent itemsets $pT_{n-1}^{\sigma}$  for the set of items $I=\{1,\ldots,n-1\}$. When
the item $(n)$ is introduced, the tree $pT_{n}^{\sigma}$  of all the frequent itemsets
for the set of items $I=\{1,\ldots,n-1,n\}$ is obtained 
 only with the use of $(n)$ and $pT_{n-1}^{\sigma}$ (\emph{see} Proposition~\ref{buildfreq}).
More precisely, this step $(n)$ is the application of the $\FIMTrav$
function to $(n)$ and $pT_{n-1}^{\sigma}$. Of course, for the association
rule mining, the updating is the application of the  $\ARMTrav$  function.

\subsubsection{Removing the first item}

Suppose that in the current step the set of items is $I=\{r,\ldots,n\}$.
Let $pT_{r,n}^{\sigma}$ be the frequent prefix tree based on $I$.
Moreover, assume that $(r)$ is frequent, so that $(r)$ is in $pT_{r,n}^{\sigma}$.
 We want to remove $(r)$
and all the frequent itemsets containing $(r)$. From the construction
of $pT_{r,n}^{\sigma}$, we see that $(r)$ is the most right child of the
root $(\emptyset)$ and all the frequent itemsets containing $(r)$
are in the subtree  rooted in the node $(r)$. We have just to remove this subtree
in $pT_{r,n}^{\sigma}$.  Of course for each deleted
itemset $C$, all the rules of type $R(A,C)$ are also deleted.

\subsection{Application: Moving FIM and Moving ARM}

In many applications, it is interesting to have the successive trees of the frequent itemsets based on a fixed number $q$ of frequent items. The  Moving FIM algorithm (MFIM) responds to this concern. The items $(i)$ of the database arrive as they go. In a first step, using the new item if it is frequent, the algorithm builds the tree of the frequent itemsets based on the first $q$ frequent items. In the second step, if the new item is frequent, the algorithm performs two tasks. In the first task it deletes all the frequent itemsets that contain the  most right child of $(\emptyset)$, that is, the 1-itemset which has spent the most time in the tree. In the second task it adds the frequent itemsets that contain the new frequent item.  
The  algorithm stops once it has performed $Q$ frequent item replacements. In the algorithm the index $i$ lists the items which successively enter the database and the index $k$ lists those which are frequent. 

Another interpretation of the Moving FIM application  is to consider only the set of $q+Q$ frequent items  $\{1,\ldots,q+Q\}$, and to see it as a moving basis of $q$   frequent items: for each integer $k$, $1\leq k\leq Q+1$,
we want to find the set $T_{k,q}^{\sigma}$ of frequent itemsets based on the set of $q$ items $\{k,\ldots,k+q-1\}$.\\

For any planar
tree $T$ with root $(\emptyset)$, we denote by $\DEL(T)$ the tree
obtained from $T$ by deleting the most right child of $(\emptyset)$
and its subtree.

\paragraph{\textit{MFIM Algorithm }}

\begin{itemize}
\item[1-] 
$k=0$, $T_{0,q}^{\sigma}=\{(\emptyset )\}$, $i=0$
 \item[2-] \textbf{while} $k\leq q$, \textbf{do}\\
 $i=i+1$\\
 \textbf{if} $(i)$ is frequent, \textbf{then}\\
 $k=k+1$\\
 $T_{k,q}^{\sigma}=\FIMTrav(T_{k-1,q}^{\sigma},(k),\sigma)$\\
 \textbf{end if} \\
\textbf{end while} 
 \item[3-] \textbf{while} $k\leq q+Q$, \textbf{do}\\ 
 $i=i+1$\\
 \textbf{if} $(i)$ is frequent, \textbf{then}\\
 $k=k+1$\\
 $T=\DEL(T_{k-1,q}^{\sigma})$\\
 $T_{k,q}^{\sigma}=\FIMTrav(T,(k),\sigma)$ \\
 \textbf{end if} \\
\textbf{end while} 

\end{itemize}

If,  for each integer $k$,  we want to find  $T_{k,q}^{\sigma}$  and, for each node $C$ of $T_{k,q}^{\sigma}$,  the set  $R(C,\tau)$,  we can use  the following Moving ARM algorithm (MARM).

\paragraph{\textit{MARM Algorithm }}
 In {\textit{MFIM Algorithm}}, replace the $\FIMTrav$ function by the $\ARMTrav$ function.

\section{Experimental Study}

\label{section comparaison} 

We conducted an experimental study on the recursion properties of PrefRec and compared it to other algorithms on computation time. For our experiments, we chose Eclat, LCM, Fp-Growth and Apriori which are among the most used and efficient algorithms for finding frequent itemsets.

 For these  algorithms, we have used the implementations of Borgelt  which can be found in  \url{https://borgelt.net/software.html}. Note that Borgelt's implementation of Eclat and LCM is in fact the implementation of an optimized mixture of Eclat and LCM, denoted Eclat/LCM. This mixture results in a more efficient program than the known versions of Eclat alone or LCM alone. We therefore chose to use Eclat/LCM in our study rather than  older versions of Eclat and LCM.  The mixture Eclat/LCM gives the possibility to select certain parameters which can affect the runtime. For all the datasets processed, we left the auto-selection of parameters which, in our case, turned out to choose the best function for almost all cases.

In order to run PrefRec, we have developed our own  C++ implementation available at  \url{https://github.com/LouisRaimbault/PrefRec}.
  The compilation of our functions  is performed with the  O3 parameters of the gcc compiler. It also includes some usual pragma directives and therefore the effectiveness may vary by machine, mainly using function to calculate the number of 1s in a bitfield. 
  The experiments are realized on 'Ubuntu 20.01 1 LTS' with characteristics: Intel Core i7-8650U CPU 1.95GHz * 8, operating system  64 bits, RAM capacity 16 GB (2*8), RAM speed 2400 Mhz.\\
Our first objective is to compare the four algorithms on the computation time necessary for the extraction of frequent itemsets, which corresponds to almost the entire time of extraction of the association rules. This study, presented in Section \ref{NRA section}, does not concern recursion properties: the databases used are fixed. 
 Let us mention that the usual known non-recursive algorithms, in order to reduce the computation time, do some actions on the database before launching the algorithm: the non frequent items are deleted, and the others are reordered according to their support. In this comparative study, the item base is fixed and we do not consider the recursion properties. So with PrefRec also, we first remove the non frequent items but the algorithm is applied without reordering the database.  However, this preprocessing of the database is not taken into account in the computation time.\\
Our second purpose is to present in Section \ref{RA section}, some recursive updating applications with PrefRec. We will present the application which consists in adding new variables to the database, as well as the Moving FIM application. Of course, in this experimental part, since the database is not fixed, we cannot preprocess it. We need to test the items as they arrive.  \\

Among the datasets we consider, $Chess$, $Accident$, $Pumsb$ and $T40I10D100K$ are taken from \url{http://fimi.uantwerpen.be/data/}.  All the others are available at  \url{https://github.com/LouisRaimbault/PrefRec}.

\begin{subsection}{Comparison with non-recursive algorithms}
\label{NRA section}

This section concerns non-recursive applications. We compare the four algorithms on the computation time. The execution time of an algorithm depends on several parameters among which we can quote: the number of transactions $N$, the number of items $n$, the supports of the items, the transaction sizes, the minSup, the number of frequent items, the number of frequent itemsets.\\  
We compare the execution times according to the minSup. We used 10 datasets, each with 7 values of minSup.\\
 In the first part, we consider four well-known datasets in data mining, namely $Chess$, $Accident$, $Pumsb$ and $T40I10D100K$. These four datasets have different characteristics. The number of items and the number of transactions are respectively 75 and 3,196 for $Chess$,  468 and 340,183 for $Accident$, 2,113 and 49,046 for $Pumsb$, 942 and 100,000 for $T40I10D100K$.\\
 In the second part we consider six synthetic datasets; two are built with independent variables and the four others are built with time series. For these six datasets, the number of transactions $N$ is 100,000. For the number of items $n$, we used the values 1000, 4000 and 6000.\\
It turns out that the results and conclusions of both parts are different.\\ 
On each graph, for a sake of representation, we do not present points of an algorithms if its runtime values are too slow compared to the others.

\begin{subsubsection}{Chess, Accident, Pumsb and T40I10D100K}
\label{Chess}

In Table \ref{table 1 simul} we give, for the  four datasets  $Chess$, $Accident$, $Pumsb$ and $T40I10D100K$, the number of frequent items (line called items) and the number of frequent itemsets (line with the dataset name) according of the minSup values. In Figure \ref{figure 1 simul} we present the execution times to make these extractions. On the $x$-axis it is minSup and on the $y$-axis it is time in seconds.

\begin {table}[h!]
\centering
\begin{tabular}{|l|c|c|c|c|c|c|c|}
  \hline
    minSup  &0.25 & 0.3 & 0.35 & 0.4 & 0.45& 0.5& 0.55 \\
  \hline
    items  &51 & 50 & 45 & 40 & 39& 37& 35 \\
  \hline
  $Chess$    & $99\ \!022\ \!533$ & $37\ \!282\ \!962$ & $15\ \!108\ \!722$ & $6\ \!439\ \!702$ & $2\ \!832\ \!777$ & $1\ \!272\ \!923$& $574\ \!998$  \\
   \hline
   minSup & 0.1 & 0.15 & 0.20 & 0.25 & 0.30 & 0.35 & 0.40 \\
   \hline
    items  &75 & 60 & 48 & 38 & 32 & 31 & 30 \\
  \hline
  $Accident$ & $10\ \!691\ \!549$ & $2\ \!676\ \!381$ & $889\ \!883$ & $346\ \!525$ & $149\ \!545$ & $68\ \!222$ & $32\ \!528$ \\
   \hline
    minSup  &0.55 & 0.60 & 0.65 & 0.70 & 0.75& 0.80 & 0.85 \\
    \hline
    items  &44 & 39 & 37 & 34 & 27& 25 & 24 \\
  \hline
  $Pumsb$    & $48\ \!790\ \!117$ & $19\ \!529\ \!991$ & $8\ \!094\ \!688$ & $2\ \!698\ \!264$ & $672\ \!390$ & $142\ \!156$& $20\ \!527$ \\
    \hline
   minSup &0.0015 & 0.002 & 0.0025 & 0.003 & 0.0035& 0.004& 0.0045 \\
   \hline
    items  &915 & 906 & 894 & 887 & 872& 862 & 848 \\
  \hline
 $T40I10D100K$  & $16\ \!917\ \!536$ & $12\ \!447\ \!895$ & $7\ \!617\ \!884$ & $5\ \!058\ \!312$ & $3\ \!439\ \!691$ & $1\ \!973\ \!395$& $1\ \!484\ \!248$ \\
 \hline

\end{tabular}
\caption{ {Number of frequent items and frequent itemsets according to the minSup value for the datasets $Chess$, $Accident$, $Pumsb$ and $T40I10D100K$}}
\label{table 1 simul}
\end{table}

\begin{figure}[H]
\vspace*{0.2cm}
\begin{tikzpicture}[scale=1]
\begin{scope}[xshift=1.5cm]
\begin{axis}[enlarge x limits=false, try min ticks = {7},
legend entries = {Eclat/LCM,Fp-Growth,PrefRec,Apriori},
legend style={at={(1.20,1.5)},anchor=north},
legend columns=4,
height = 4.75 cm,
width = 6.1 cm,
axis x line = bottom,
axis y line = left,
grid = major,
 title = {Chess},
xlabel = {minSup$\times 10$},
]
\addplot  coordinates {(2.5,1.5) (3,0.73) (3.5,0.37) (4,0.19) (4.5,0.1) (5,0.06) (5.5,0.03) };
\addplot +[mark=x,mark options = {xscale = 1.5, yscale = 1.5 },color = red]  coordinates {(2.5,0.53) (3,0.25) (3.5,0.12) (4,0.06) (4.5,0.03) (5,0.02) (5.5,0.01) };
\addplot +[mark=o,mark options={xscale = 1.5, yscale = 1.5}, color = black]  coordinates {(3,3.35) (3.5,1.28) (4,0.54) (4.5,0.25) (5,0.1) (5.5,0.044329) };
\addplot +[mark=*,mark options = {xscale = 1.5, yscale = 1.5 }, color = brown ]  coordinates {(5,6) (5.5,2.32) };

\end{axis}
\end{scope}
\begin{scope}[xshift = 7cm]
\begin{axis}[enlarge x limits=false, try min ticks = {7},
height = 4.75 cm,
width = 6.1 cm,
axis x line = bottom,
axis y line = left,
grid = major,
 title = {Accident},
xlabel = {minSup$\times 10$},
]

\addplot  coordinates {(1,10.95) (1.5,6.73) (2,5.15) (2.5,3.19) (3,1.50) (3.5,1.30) (4,0.97) };
\addplot +[mark=x,mark options = {xscale = 1.5, yscale = 1.5 },color = red]  coordinates {(1,2.82) (1.5,1.54) (2,0.95) (2.5,0.50) (3,0.24) (3.5,0.18) (4,0.15) };
\addplot +[mark=o,mark options={xscale = 1.5, yscale = 1.5}, color = black]  coordinates { (1.5,7.91) (2,2.89) (2.5,1.25) (3,0.61) (3.5,0.34) (4,0.20) };
\addplot +[mark=*,mark options = {xscale = 1.5, yscale = 1.5 }, color = brown ]  coordinates {};

\end{axis}
\end{scope}
\end{tikzpicture}
\begin{tikzpicture}[scale=1]
\begin{scope}[xshift=1.5cm]
\begin{axis}[enlarge x limits=false, try min ticks = {7},
legend entries = {Eclat/LCM,Fp-Growth,PrefRec,Apriori},
legend style={at={(1.20,1.5)},anchor=north},
legend columns=4,
height = 4.75 cm,
width = 6.1 cm,
axis x line = bottom,
axis y line = left,
grid = major,
 title = {Pumsb},
xlabel = {minSup$\times 10$},
]
\addplot  coordinates {(5.5,1.33) (6,0.51) (6.5,0.25) (7,0.13) (7.5,0.06) (8,0.02) (8.5,0.01)};
\addplot +[mark=x,mark options = {xscale = 1.5, yscale = 1.5 },color = red]  coordinates {(5.5,0.30) (6,0.12) (6.5,0.06) (7,0.03) (7.5,0.01) (8,0.01) (8.5,0.01) };
\addplot +[mark=o,mark options={xscale = 1.5, yscale = 1.5}, color = black]  coordinates { (6.5,3.68) (7,1.27) (7.5,0.34) (8,0.09) (8.5,0.02) };
\addplot +[mark=*,mark options = {xscale = 1.5, yscale = 1.5 }, color = brown ]  coordinates {(7,2.58) (7.5,0.62) (8,0.19) (8.5,0.05) };

\end{axis}
\end{scope}
\begin{scope}[xshift = 7cm]
\begin{axis}[enlarge x limits=false, try min ticks = {7},
height = 4.75 cm,
width = 6.1 cm,
axis x line = bottom,
axis y line = left,
grid = major,
 title = {T40I10D100K},
xlabel = {minSup$\times 10^{3}$},
]
\addplot  coordinates {(1.5,7.6) (2,5.8) (2.5,4.68) (3,3.90) (3.5,3.33) (4,2.91) (4.5,2.59) };
\addplot +[mark=x,mark options = {xscale = 1.5, yscale = 1.5 },color = red]  coordinates {(1.5,11.77) (2,9.67) (2.5,8.30) (3,7.45) (3.5,6.88) (4,6.43) (4.5,5.93) };
\addplot +[mark=o,mark options={xscale = 1.5, yscale = 1.5}, color = black]  coordinates {(1.5,6.67) (2,4.37) (2.5,2.99) (3,2.24) (3.5,1.74) (4,1.39) (4.5,1.20) };
\addplot +[mark=*,mark options = {xscale = 1.5, yscale = 1.5 }, color = brown ]  coordinates {};

\end{axis}
\end{scope}
\end{tikzpicture}
  \caption{Runtime for Chess, Accident, Pumsb and T40I10D100K}
  \label{figure 1 simul}
\end{figure}
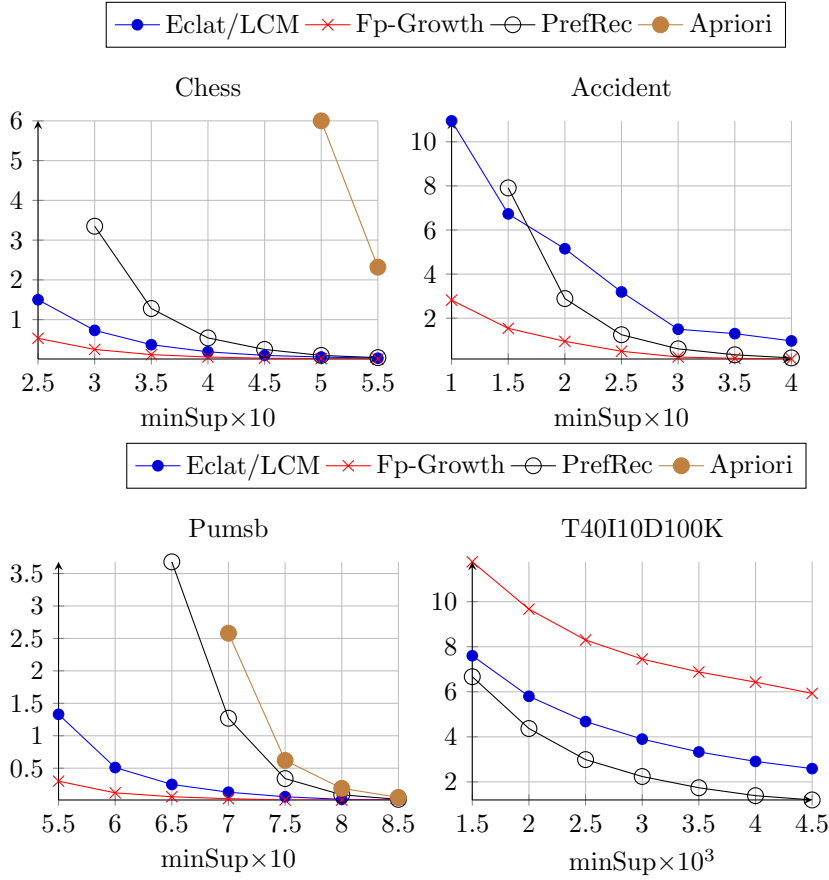

In view of the  results, we first notice that Apriori has many difficulties and is clearly very inefficient compared to the other algorithms. It is clear that Fp-Growth performs better than all the others for the $Chess$, $Accident$ and $Pumbs$ datasets.  Eclat/LCM performs better than PrefRec for $Chess$ and $Pumbs$. But it does not do much better for the $Accident$ dataset where it seems to lose some effectiveness. This is probably due to the fact that the number of transactions $N$ is large in the $Accident$ database.\\
The results are different with the $T40I10D100K$ dataset. In this case, Eclat/LCM and PrefRec are the best with a slight advantage for PrefRec. Both algorithms work slightly better than Fp-Growth which seems to have some difficulties with the $T40I10D100K$ dataset.\\
 The main difference between $T40I10D100K$ and the other datasets is the number of frequent 1-items (see Table \ref{table 1 simul}). For $T40I10D100K$ there are more than 860 frequent 1-items, while for the others the number of frequent 1-items is less than 76. 
It therefore seems that Fp-Growth is suitable for the case where the number of frequent 1-items is small and suffers a lot when this number increases. Eclat/LCM and PrefRec seem more resistant to cases where the number of frequent 1-items is large. 
This will be confirmed with the experiment on synthetic datasets.

\end{subsubsection}

\begin{subsubsection}{Synthetic datasets}
\label{Synthetic}

We now present the comparisons on synthetic data sets.
For all the data sets we have simulated, the number of individuals is $N = 100,000$. Four datasets have a number of items $n = 1,000$.
Two of these four data sets are constructed using a sequence of independent variables and the other two using a time series.\\
To confirm the importance of the number of frequent 1-items, and thus also that of the parameter $n$, we added two data sets with respectively $n = 4,000$ and $n = 6,000$. These added datasets are also obtained using a time series.\\
In each dataset, the transaction sizes remain quite close.
In the names of the datasets, $n$ is the number of items, $N$ is the number of individuals in thousands and $Me$ is the averaged number of items per transaction.\\

The two independant datasets are generated as follows. For each of the $n$ variables, we first choose randomly a value $p$ in some given interval $J\subset[0,1]$ and then we generate an $N$-sample of the Bernoulli distribution ${\cal B}(p)$. The chosen intervals $J$ are given in Tables~\ref{table 2 simul} and~\ref{table 3 simul}. For example, line 1 in Table~\ref{table 2 simul} says that for $30\%$ of the $n$ variables, the parameter $p$ of each variable has been chosen uniformly between 0.005 and 0.08.\\
The first independent data set, called $In1000N100Me103$, is generated according to Table~\ref{table 2 simul}, the second called $In1000N100Me163$, according to Table~\ref{table 3 simul}.

\begin {table}[h!]
\centering
\begin{tabular}{|c|c|}
    \hline
    Number of variables &  ${\cal B}(p)$  \\
    \hline
    30 \% & 0.005 $\leq p \leq$ 0.08 \\
  30 \% & 0.08 $\leq p \leq$ 0.12 \\
  30 \% & 0.12 $\leq p \leq$ 0.15 \\
  10 \% & 0.15 $\leq p \leq$ 0.25\\
    \hline
\end{tabular}
\caption{ { The Bernoulli variables in the dataset $In1000N100Me103$}}
\label{table 2 simul}
\end {table}

\begin {table}[h!]
\centering
\begin{tabular}{|c|c|}
    \hline
    Number of variables &  ${\cal B}(p)$ \\
    \hline
    20 \% & 0.05 $\leq p \leq$ 0.13 \\
    35 \% & 0.13 $\leq p \leq$ 0.16 \\
    35 \% & 0.16 $\leq p \leq$ 0.20 \\
    10 \% & 0.20 $\leq p \leq$ 0.40\\
    \hline
\end{tabular}
\caption{ { The Bernoulli variables in the dataset In1000N100Me163}}
\label{table 3 simul}
\end {table}

Dependent datasets are generated as follows. We consider an autogressive process, $AR(3)$,  defined by  $Z^{(t)} = 1.15 Z^{(t-1)} -0.06 Z^{(t-2)} -0.1485 Z^{(t-3)} + \epsilon^{(t)} $ where $\epsilon^{(t)} $ is a Gaussian white noise with a variance equal to 0.5. 
 The time index $t$ takes its values in $\{1,\ldots ,n\}$.
We fix a threshold $s$ and set $X^{(t)}$ equal to~1 if $Z^{(t)}>s$ and to~0 otherwise. We generate $N=100,000$ independent autoregressive processes 
$(Z^{(t)})_{i}$, $1 \leq i \leq N$, leading to a sample
 $(X^{(t)})_{i}$, $1 \leq i \leq N$.  We consider two values of tresholds, $s=2.3$  and $s=1.7$.\\
For $n=1,000$, we generate the datasets $ARn1000N100Me103$ with $s=2.3$ and $ARn1000N100Me163$ with $s=1.7$. The tresholds $s$ were chosen such that for $n=1,000$, the dependent synthetic data sets have the same $Me$ as the independent synthetic data sets.\\
With $n=4,000$ and $s=1.7$, we generate $ARn4000N100Me658$ and with $n=6,000$ and $s=1.7$ we generate $ARn6000N100Me987$.\\

Table \ref{table 4 simul} below is organized like Table \ref{table 1 simul}. It gives for the four  datasets with $n=1000$, the number of frequent items and the number of frequent itemsets according of the minSup values.  Figure \ref{figure 2 simul}  presents the execution times to make these extractions.

\begin {table}[h!]
\centering
\begin{tabular}{|l|c|c|c|c|c|c|c|}
  \hline
    minSup & 0.004&0.005&0.006&0.007&0.008&0.009&0.01 \\
  \hline
  items & 1000 & 1000 & 995 & 989 & 986 & 980 & 977 \\
  \hline
  $In1000N100Me103$    & $4\ \!009\ \!246$ & $1\ \!698\ \!133$ & $940\ \!501$ & $567\ \!756$ & $382\ \!110$ & $305\ \!383$& $263\ \!334$  \\
    \hline
    minSup &0.015&0.02&0.025&0.03&0.035&0.04&0.045 \\
   \hline
   items & 1000 & 1000 & 1000 & 1000 & 1000 & 1000 & 1000 \\
  \hline
  $In1000N100Me163$ & $4\ \!212\ \!703$ & $1\ \!493\ \!372$ & $662\ \!189$ & $339\ \!070$ & $187\ \!319$ & $120\ \!438$ & $80\ \!752$ \\
   \hline
    minSup & 0.01&0.015&0.02&0.025&0.03&0.035&0.040 \\
   \hline
   items & 1000 & 1000 & 1000 & 1000 & 1000 & 1000 & 1000 \\
  \hline
    $ARn1000N100Me103$  & $9\ \!585\ \!055$ & $1\ \!596\ \!769$ & $493\ \!628$ & $187\ \!578$ & $88\ \!097$ & $47\ \!300$ & $29\ \!748$ \\
    \hline
   minSup &0.025&0.03&0.035&0.04&0.045&0.05&0.055 \\
   \hline
   items & 1000 & 1000 & 1000 & 1000 & 1000 & 1000 & 1000 \\
  \hline
 $ARn1000N100Me163$   & $7\ \!860\ \!411$ & $2\ \!878\ \!483$ & $1\ \!303\ \!059$ & $650\ \!431$ & $348\ \!069$ & $203\ \!940$ & $132\ \!006$\\
 \hline
 
\end{tabular}
\caption{ {Number of frequent items and frequent itemsets according to the minSup value for synthetic datasets with $n=1,000$}}
\label{table 4 simul}
\end{table}

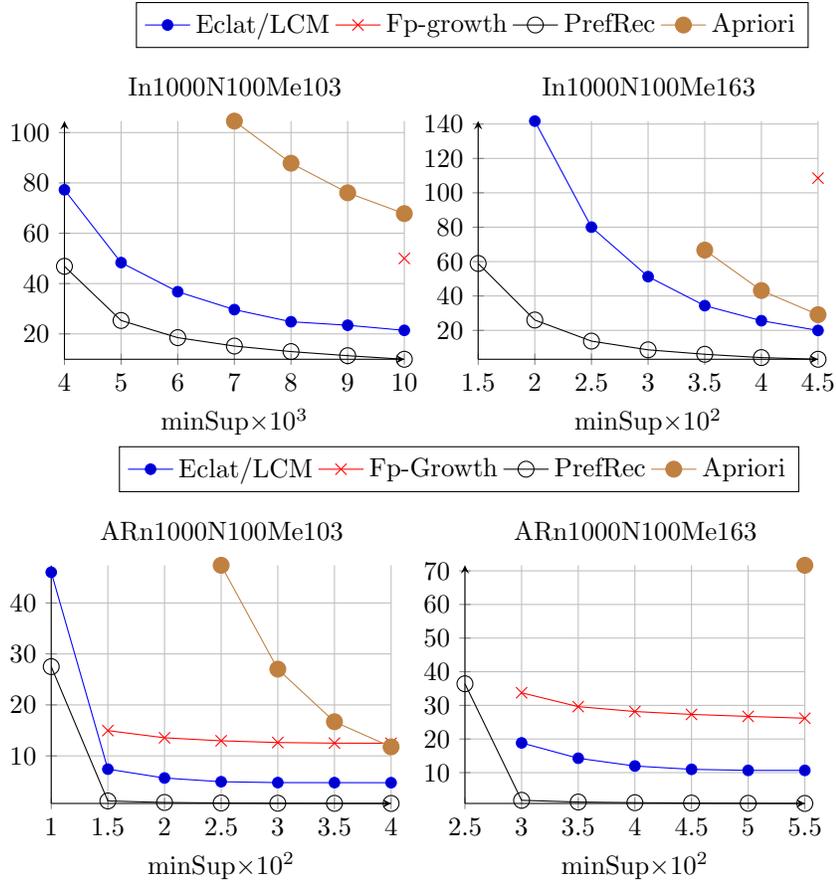
\begin{figure}[H]
\vspace*{0.2cm}
\begin{tikzpicture}[scale=1]
\begin{scope}[xshift=1.5cm]
\begin{axis}[enlarge x limits=false, try min ticks = {7},
legend entries = {Eclat/LCM,Fp-growth,PrefRec,Apriori},
legend style={at={(1.20,1.5)},anchor=north},
legend columns=4,
height = 4.75 cm,
width = 6.1 cm,
axis x line = bottom,
axis y line = left,
grid = major,
 title = {In1000N100Me103},
xlabel = {minSup$\times 10^{3}$},
]
\addplot  coordinates {(4,77.26) (5,48.33) (6,36.76) (7,29.65) (8,24.83) (9,23.44) (10,21.41) };
\addplot +[mark=x,mark options = {xscale = 1.5, yscale = 1.5 },color = red]  coordinates {(10,50) };
\addplot +[mark=o,mark options={xscale = 1.5, yscale = 1.5}, color = black]  coordinates {(4,46.85) (5,25.33) (6,18.51) (7,15.17) (8,12.97) (9,11.34) (10,9.92) };
\addplot +[mark=*,mark options = {xscale = 1.5, yscale = 1.5 }, color = brown ]  coordinates {(7,104.56) (8,87.83) (9,76.07) (10,67.8) };

\end{axis}
\end{scope}
\begin{scope}[xshift = 7cm]
\begin{axis}[enlarge x limits=false, try min ticks = {7},
height = 4.75 cm,
width = 6.1 cm,
axis x line = bottom,
axis y line = left,
grid = major,
 title = {In1000N100Me163},
xlabel = {minSup$\times 10^{2}$},
]
\addplot  coordinates {(2,141.7) (2.5,80.04) (3,51.25) (3.5,34.37) (4,25.62) (4.5,19.96) };
\addplot +[mark=x,mark options = {xscale = 1.5, yscale = 1.5 },color = red]  coordinates {(4.5,108.54) };
\addplot +[mark=o,mark options={xscale = 1.5, yscale = 1.5}, color = black]  coordinates {(1.5,58.91) (2,26.02) (2.5,13.75) (3,8.67) (3.5,6.09) (4,4.16) (4.5,3.19) };
\addplot +[mark=*,mark options = {xscale = 1.5, yscale = 1.5 }, color = brown ]  coordinates {(3.5,66.68) (4,43.17) (4.5,29.13) };

\end{axis}
\end{scope}
\end{tikzpicture}
\begin{tikzpicture}[scale=1]
\begin{scope}[xshift=1.5cm]
\begin{axis}[enlarge x limits=false, try min ticks = {7},
legend entries = {Eclat/LCM,Fp-Growth,PrefRec,Apriori},
legend style={at={(1.20,1.5)},anchor=north},
legend columns=4,
height = 4.75 cm,
width = 6.1 cm,
axis x line = bottom,
axis y line = left,
grid = major,
 title = {ARn1000N100Me103},
xlabel = {minSup$\times 10^{2}$},
]
\addplot  coordinates {(1,46.02) (1.5,7.4) (2,5.66) (2.5,4.93) (3,4.76) (3.5,4.75) (4,4.74) };
\addplot +[mark=x,mark options = {xscale = 1.5, yscale = 1.5 },color = red]  coordinates {(1,nan) (1.5,14.96) (2,13.55) (2.5,12.96) (3,12.62) (3.5,12.5) (4,12.48) };
\addplot +[mark=o,mark options={xscale = 1.5, yscale = 1.5}, color = black]  coordinates {(1,27.51) (1.5,1.16) (2,0.86) (2.5,0.76) (3,0.71) (3.5,0.69) (4,0.67) };
\addplot +[mark=*,mark options = {xscale = 1.5, yscale = 1.5 }, color = brown ]  coordinates {(2.5,47.39) (3,27.02) (3.5,16.7) (4,11.77) };

\end{axis}
\end{scope}
\begin{scope}[xshift = 7cm]
\begin{axis}[enlarge x limits=false, try min ticks = {7},
height = 4.75 cm,
width = 6.1 cm,
axis x line = bottom,
axis y line = left,
grid = major,
 title = {ARn1000N100Me163},
xlabel = {minSup$\times 10^{2}$},
]
\addplot  coordinates {(3,18.84) (3.5,14.28) (4,11.97) (4.5,10.98) (5,10.67) (5.5,10.67) };
\addplot +[mark=x,mark options = {xscale = 1.5, yscale = 1.5 },color = red]  coordinates {(3,33.75) (3.5,29.62) (4,28.17) (4.5,27.31) (5,26.72) (5.5,26.18) };
\addplot +[mark=o,mark options={xscale = 1.5, yscale = 1.5}, color = black]  coordinates {(2.5,36.423) (3,1.77) (3.5,1.26) (4,1.07) (4.5,0.95) (5,0.88) (5.5,0.84) };
\addplot +[mark=*,mark options = {xscale = 1.5, yscale = 1.5 }, color = brown ]  coordinates {(5.5,71.65) };

\end{axis}
\end{scope}
\end{tikzpicture}
  \caption{Running time in seconds for synthetic datasets with $n=1,000$}
  \label{figure 2 simul}
  
\end{figure}

In view of these results with $n = 1,000$, it is clearly seen that Apriori and Fp-Growth  are very inefficient compared to Eclat/LCM and PrefRec. We also see that PrefRec is slightly more efficient than Eclat/LCM in these four datasets.\\

Let us go further and now consider the case where $n$ is greater.\\ 

The results for the synthetic datasets with $n=4,000$ and $n=6,000$ are given  in Table \ref{table 5 simul} and Figure \ref{figure 3 simul} below. In Figure \ref{figure 3 simul}, only the times taken by PrefRec and Eclat/LCM are shown. The execution times of Apriori and Fp-Growth are too long to be represented.

\begin {table}[h!]
\centering
\begin{tabular}{|l|c|c|c|c|c|c|c|}
  \hline
    minSup  &0.08 & 0.03 & 0.035 & 0.04 & 0.045& 0.05& 0.055 \\
  \hline
    items  &4000 & 4000 & 4000 & 4000 & 4000 & 4000 & 4000 \\
  \hline
  $ARn4000N100Me658$  & $16\ \!769\ \!590$ & $11\ \!517\ \!755$ & $5\ \!214\ \!935$ & $2\ \!604\ \!917$ & $1\ \!394\ \!004$ &  $815\ \!199$ & $528\ \!017$  \\
  \hline
    minSup  &0.08 & 0.03 & 0.035 & 0.04 & 0.045& 0.05& 0.055 \\
  \hline
    items  &6000 & 6000 & 6000 & 6000 & 6000 & 6000 & 6000 \\
  \hline
  $ARn6000N100Me987$ & $25868628$ & $17487275$ & $7899798$ & $3940726$ & $2110140$ & $1233128$ & $796669$ \\
   \hline

\end{tabular}
\caption{ {Number of frequent items and frequent itemsets according to the minSup value for synthetic datasets with large $n$}}
\label{table 5 simul}
\end{table}

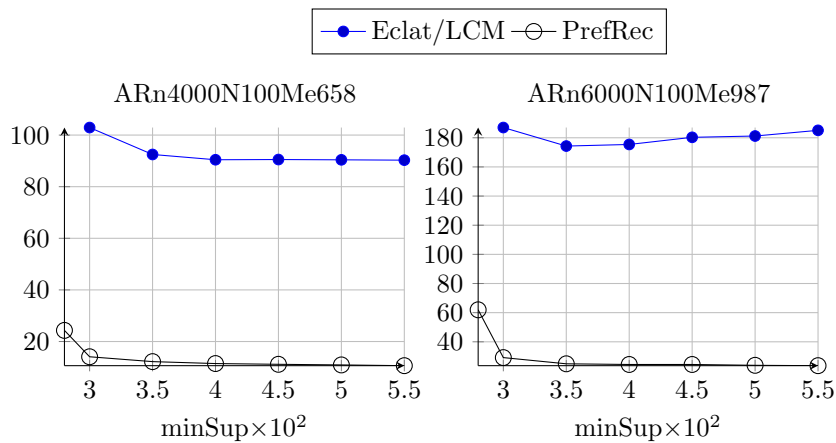
\begin{figure}[H]
\vspace*{0.2cm}
\begin{tikzpicture}[scale=1]
\begin{scope}[xshift=1.5cm]
\begin{axis}[enlarge x limits=false, try min ticks = {7},
legend entries = {Eclat/LCM,PrefRec},
legend style={at={(1.25,1.5)},anchor=north},
legend columns=4,
height = 4.75 cm,
width = 6.1 cm,
axis x line = bottom,
axis y line = left,
grid = major,
 title = {ARn4000N100Me658},
xlabel = {minSup$\times 10^{2}$},
]
\addplot  coordinates { (3,102.92) (3.5,92.48) (4,90.44) (4.5,90.52) (5,90.41) (5.5,90.28) };
\addplot +[mark=o,mark options={xscale = 1.5, yscale = 1.5}, color = black]  coordinates {(2.8,24.31) (3,14.09) (3.5,12.21) (4,11.46) (4.5,11.17) (5,10.93) (5.5,10.64) };

\end{axis}
\end{scope}
\begin{scope}[xshift = 7cm]
\begin{axis}[enlarge x limits=false, try min ticks = {7},
height = 4.75 cm,
width = 6.1 cm,
axis x line = bottom,
axis y line = left,
grid = major,
 title = {ARn6000N100Me987},
xlabel = {minSup$\times 10^{2}$},
]
\addplot  coordinates { (3,187.03) (3.5,174.31) (4,175.41) (4.5,180.31) (5,181.20) (5.5,185.10) };
\addplot +[mark=o,mark options={xscale = 1.5, yscale = 1.5}, color = black]  coordinates {(2.8,61.95) (3,29.16) (3.5,24.90) (4,24.44) (4.5,24.44) (5,23.87) (5.5,23.62) };

\end{axis}
\end{scope}
\end{tikzpicture}
  \caption{Running time in seconds for synthetic datasets with large $n$}
  \label{figure 3 simul}
  
\end{figure}

In these last two experiments, PrefRec seems, once again, a little more efficient than Eclat/LCM. This confirms the results of the previous four experiments.\\

For the 10 datasets we used, we can conclude that PrefRec is more efficient than Apriori in all cases and that in general it supports the comparison very well with the other algorithms. In the case when the number of frequent items is very low, it has a little more difficulty than Fp-Growth and Eclat/LCM. But as soon as this number is higher it becomes more efficient than Fp-Growth and quite comparable with Eclat/LCM.
Note also that PrefRec seems particularly effective in the case of time series. This is probably because it is recursive.
 
\end{subsubsection}

\end{subsection}

\subsection{Applications of recursion properties}

\label{RA section}

In this part, we present applications that can be made thanks to the recursion properties of PrefRec. These applications show the important and innovative properties of PrefRec. They cannot be performed with the other algorithms. 
 A major advantage of the recursion properties is that all the variables of the database do not need to be definitively entered before the start of the algorithm: the data can be processed as and when they arrive. This is particularly useful in the case of time series but not only. There is therefore no pre-processing of the dataset.

\subsubsection{Adding Variables}

We present here the results of the experiment consisting in adding new variables to a database. An initial database is given. Then, new variables are introduced into the model; we measure the time needed to update the list of frequent itemsets.

We consider the independent case and the dependent case.
In both cases, the notation $re$ indicates the number of variables introduced into the model.\\
 
In the independent case, the initial bases considered are $In1000N100Me103$ and $In1000N100Me163$. When we introduce a single variable, ($re1$), it is a Bernoulli variable with parameter $p$ chosen randomly in a given interval. As example the title $re1[0.08:0.12]$ means that a Bernoulli variable with a parameter $p$, chosen randomly in $[0.08 , 0.12]$, was added a posteriori to the 1000 variables of the model.

The title $re10$ (resp. $re100$) means that a group of 10 (resp. 100) variables has been added to the initial database. This group is generated respecting the distributions given in Table~\ref{table 2 simul} for $In1000N100Me103$ (resp. Table~\ref{table 3 simul} for $In1000N100Me163$).\\
 
In the dependent case, the initial bases considered are $ARn1000N100Me103$ and $ARn1000N100Me163$. The added variables are obtained from the time series. In the case $re1$, the added variable is $X^{(n+1)}$. In the case $re10$ (resp. $re100$) the added variables are $X^{(n+1)},...,X^{(n+10)}$ (resp.  $X^{(n+1)},...,X^{(n+100)}$).\\    

Because they are not intended for this, Eclat/LCM, Apriori and Fp-Growth must be restarted for each added variables, unlike PrefRec which was defined and built around recursive logic. Therefore we only present the calculation times of PrefRec, the calculation times of the others algorithms being too long to be represented.

\begin{figure}[H]
\begin{tikzpicture}[scale = 1]
\begin{scope} 
\begin{axis}[enlarge x limits=false,try min ticks={7},
             xtick = {0.5,0.7,1,1.2,1.5,2,2.5},
             legend entries = {PrefRec},
             legend style = {nodes={scale=0.63, transform shape}, at = {(0.52,1)},anchor = north west},  
             height = 4.4 cm,
             width = 6.1 cm, 
             axis x line = bottom,
             axis y line = left,
             grid = major,
             title = {$re1[0.005:0.08]$},
             xlabel = {minSup value $\times 10^2$},
             ]
\addplot coordinates {(0.5,0.025) (0.7,0.015) (1,0.007) (1.2,0.005)  (1.5,0.0043) (2,0.0037) (2.5,0.0036)};  

\end{axis}
\end{scope}
\begin{scope}[xshift= 6 cm]
\begin{axis}[enlarge x limits=false,try min ticks={7},
             xtick = {0.5,0.7,1,1.2,1.5,2,2.5},
             legend entries = {PrefRec},
             legend style = {nodes={scale=0.63, transform shape}, at = {(0.52,1)},anchor = north west},   
             height = 4.4 cm,
             width = 6.1 cm, 
             axis x line = bottom,
             axis y line = left,
             grid = major,
             title = {$re1[0.08:0.12]$},
             xlabel = {minSup value $\times 10^2$},
             ]
\addplot coordinates {(0.5,0.091) (0.7,0.059) (1,0.041) (1.2,0.028)  (1.5,0.016) (2,0.009) (2.5,0.008)};  
\end{axis}
\end{scope}
\end{tikzpicture}

\begin{tikzpicture}[scale = 1]
\begin{scope} [xshift= 0.4 cm]
\begin{axis}[enlarge x limits=false,try min ticks={7},
             xtick = {0.5,0.7,1,1.2,1.5,2,2.5},
             legend entries = {PrefRec},
             legend style = {nodes={scale=0.63, transform shape}, at = {(0.52,1)},anchor = north west},  
             height = 4.4cm,
             width = 6.1 cm, 
             axis x line = bottom,
             axis y line = left,
             grid = major,
             title = {$re1[0.12:0.15]$},
             xlabel = {minSup value $\times 10^2$},
             ]
\addplot coordinates {(0.5,0.16) (0.7,0.087) (1,0.065) (1.2,0.054)  (1.5,0.038) (2,0.017) (2.5,0.012)};  

\end{axis}
\end{scope}
\begin{scope}[xshift= 6.5 cm]
\begin{axis}[enlarge x limits=false,try min ticks={7},
             xtick = {0.5,0.7,1,1.2,1.5,2,2.5},
             legend entries = {PrefRec},
             legend style = {nodes={scale=0.63, transform shape}, at = {(0.52,1)},anchor = north west},   
             height = 4.4 cm,
             width = 6.1 cm, 
             axis x line = bottom,
             axis y line = left,
             grid = major,
             title = {$re1[0.15:0.25]$},
             xlabel = {minSup value $\times 10^2$},
             ]
\addplot coordinates {(0.5,0.361) (0.7,0.131) (1,0.094) (1.2,0.082)  (1.5,0.065) (2,0.044) (2.5,0.030)};

\end{axis}
\end{scope}
\end{tikzpicture}

\begin{tikzpicture}[scale = 1]
\begin{scope} [xshift= 1.5 cm]
\begin{axis}[enlarge x limits=false,try min ticks={7},
             xtick = {0.5,0.7,1,1.2,1.5,2,2.5},
             legend entries = {PrefRec},
             legend style = {nodes={scale=0.63, transform shape}, at = {(0.52,1)},anchor = north west},  
             height = 4.4cm,
             width = 6.1 cm, 
             axis x line = bottom,
             axis y line = left,
             grid = major,
             title = {$re10$},
             xlabel = {minSup value $\times 10^2$},
             ]
\addplot coordinates {(0.5,1.3) (0.7,0.7) (1,0.45) (1.2,0.336)  (1.5,0.228) (2,0.130) (2.5,0.101)};  
\end{axis}
\end{scope}
\begin{scope}[xshift= 8 cm]
\begin{axis}[enlarge x limits=false,try min ticks={7},
             xtick = {0.5,0.7,1,1.2,1.5,2,2.5},
             legend entries = {PrefRec},
             legend style = {nodes={scale=0.63, transform shape}, at = {(0.52,1)},anchor = north west},   
             height = 4.4cm,
             width = 6.1 cm, 
             axis x line = bottom,
             axis y line = left,
             grid = major,
             title = {$re100$},
             xlabel = {minSup value $\times 10^2$},
             ]
\addplot coordinates {(0.5,11.85) (0.7,6.75) (1,4.38) (1.2,3.31)  (1.5,2.04) (2,1.106) (2.5,1.013)};  
             
\end{axis}
\end{scope}
\end{tikzpicture}

\caption{Execution time in seconds to add variables to the database $In1000N100Me103$}
\label{fig 2.1 simul}
\end{figure}
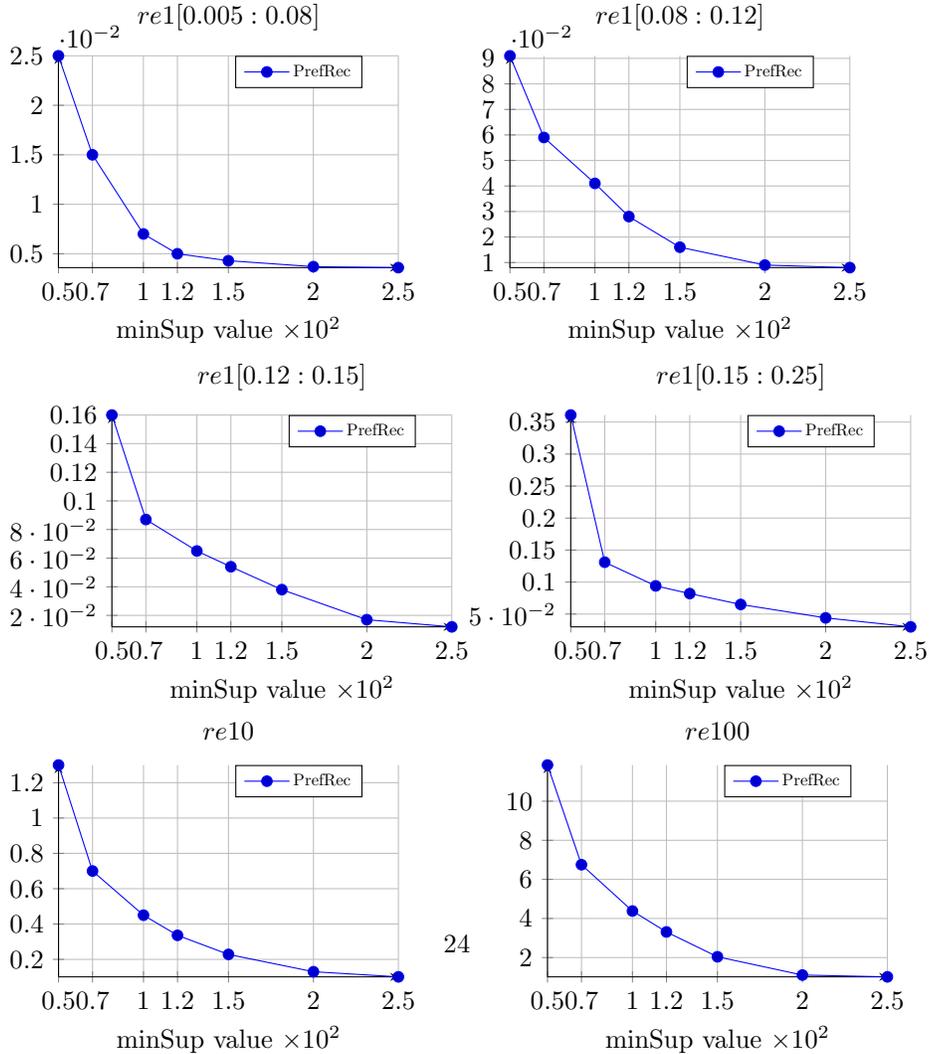

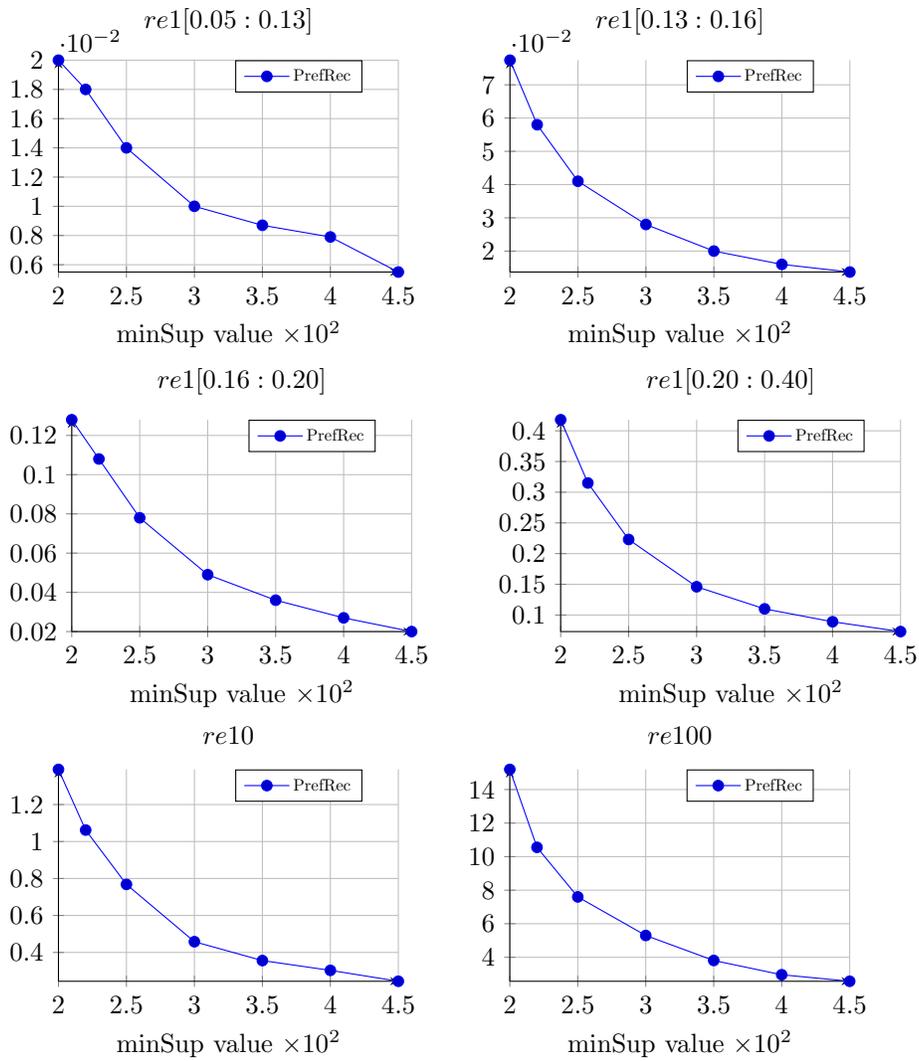
\begin{figure} [H]

\begin{tikzpicture}[scale = 1]
\begin{scope} [xshift= 0.5 cm]
\begin{axis}[enlarge x limits=false,try min ticks={7},
             legend entries = {PrefRec},
             legend style = {nodes={scale=0.63, transform shape}, at = {(0.52,1)},anchor = north west},  
             height = 4.4 cm,
             width = 6.1 cm, 
             axis x line = bottom,
             axis y line = left,
             grid = major,
             title = {$re1[0.05:0.13]$},
             xlabel = {minSup value $\times 10^2$},
             ]
\addplot coordinates {(2,0.020) (2.2,0.018) (2.5,0.014) (3,0.01)  (3.5,0.0087) (4,0.0079) (4.5,0.0055)};  

\end{axis}
\end{scope}
\begin{scope}[xshift= 6.5 cm]
\begin{axis}[enlarge x limits=false,try min ticks={7},
             legend entries = {PrefRec},
             legend style = {nodes={scale=0.63, transform shape}, at = {(0.52,1)},anchor = north west},   
             height = 4.4 cm,
             width = 6.1 cm, 
             axis x line = bottom,
             axis y line = left,
             grid = major,
             title = {$re1[0.13:0.16]$},
             xlabel = {minSup value $\times 10^2$},
             ]
\addplot coordinates {(2,0.0774) (2.2,0.058) (2.5,0.041) (3,0.028)  (3.5,0.020) (4,0.016) (4.5,0.0137)};

\end{axis}
\end{scope}
\end{tikzpicture}

\begin{tikzpicture}[scale = 1]
\begin{scope} [xshift= 1.5 cm]
\begin{axis}[enlarge x limits=false,try min ticks={7},
             legend entries = {PrefRec},
             legend style = {nodes={scale=0.63, transform shape}, at = {(0.52,1)},anchor = north west},  
             height = 4.4cm,
             width = 6.1 cm, 
             axis x line = bottom,
             axis y line = left,
              ytick = {0.02,0.04,0.06,0.08,0.1,0.12},
             yticklabels = {0.02,0.04,0.06,0.08,0.1,0.12},
             grid = major,
             title = {$re1[0.16:0.20]$},
             xlabel = {minSup value $\times 10^2$},
             ]
\addplot coordinates {(2,0.128) (2.2,0.108) (2.5,0.078) (3,0.049)  (3.5,0.036) (4,0.027) (4.5,0.02)};
  
\end{axis}
\end{scope}
\begin{scope}[xshift= 8 cm]
\begin{axis}[enlarge x limits=false,try min ticks={7},
             legend entries = {PrefRec},
             legend style = {nodes={scale=0.63, transform shape}, at = {(0.52,1)},anchor = north west},   
             height = 4.4cm,
             width = 6.1 cm, 
             axis x line = bottom,
             axis y line = left,
             grid = major,
             title = {$re1[0.20:0.40]$},
             xlabel = {minSup value $\times 10^2$},
             ]
\addplot coordinates {(2,0.418) (2.2,0.315) (2.5,0.223) (3,0.146)  (3.5,0.110) (4,0.089) (4.5,0.073)};  
             
\end{axis}
\end{scope}
\end{tikzpicture}

\begin{tikzpicture}[scale = 1]
\begin{scope} [xshift= 0.5 cm]
\begin{axis}[enlarge x limits=false,try min ticks={7},
             legend entries = {PrefRec},
             legend style = {nodes={scale=0.63, transform shape}, at = {(0.52,1)},anchor = north west},  
             height = 4.4 cm,
             width = 6.1 cm, 
             axis x line = bottom,
             axis y line = left,
             grid = major,
             title = {$re10$},
             xlabel = {minSup value $\times 10^2$},
             ]
\addplot coordinates {(2,1.39) (2.2,1.062) (2.5,0.768) (3,0.458)  (3.5,0.356) (4,0.303) (4.5,0.244)};  

\end{axis}
\end{scope}
\begin{scope}[xshift= 6.5 cm]
\begin{axis}[enlarge x limits=false,try min ticks={7},
             legend entries = {PrefRec},
             legend style = {nodes={scale=0.63, transform shape}, at = {(0.52,1)},anchor = north west},   
             height = 4.4 cm,
             width = 6.1 cm, 
             axis x line = bottom,
             axis y line = left,
             grid = major,
             title = {$re100$},
             xlabel = {minSup value $\times 10^2$},
             ]
\addplot coordinates {(2,15.20) (2.2,10.553) (2.5,7.60) (3,5.30)  (3.5,3.81) (4,2.958) (4.5,2.57)};

\end{axis}
\end{scope}
\end{tikzpicture}

\caption{Execution time in seconds to add variables to the database $In1000N100Me163$
}
\label{fig 2.2 simul}
\end{figure}

Since the results are random, we conducted 50 simulations. The results presented are averages over the 50 simulations.  
Table~\ref{table 11} (for the independent datasets) and Table~\ref{table 12} (for the dependent datasets)  indicate the number of new frequent itemsets after adding variables. Figures~\ref{fig 2.1 simul} and~\ref{fig 2.2 simul} (for the independent datasets) and Figure~\ref{fig 2.3 simul} (for the dependent datasets) give the running time as a function of minSup.\\ 

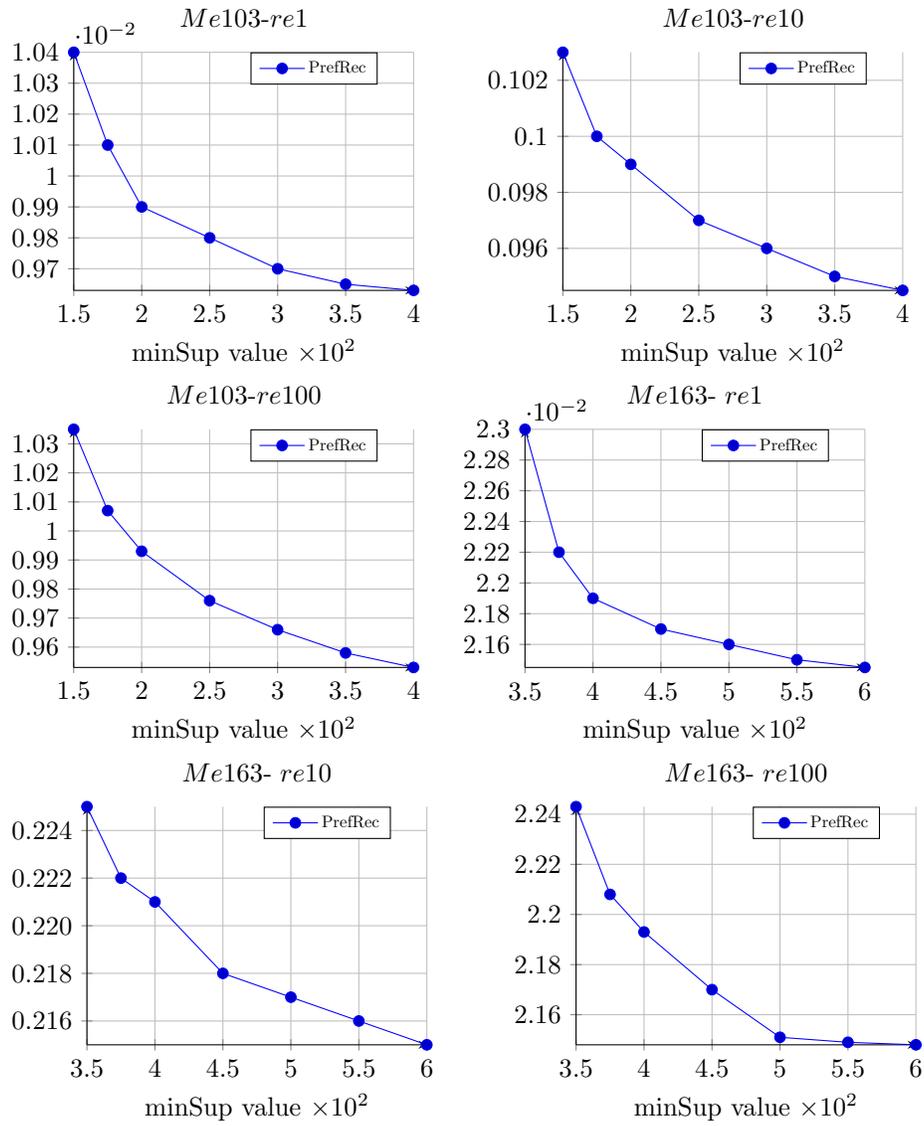
\begin{table}[h!]
\centering
\begin{tabular}{|l|c|c|c|c|c|c|c|}
  \hline
   $In1000N100Me103$ & 0.005 &0.007 &0.01& 0.012 & 0.015 & 0.02& 0.025\\
  \hline
   $re1[0.005:0.08]$ & 345 & 203 & 70 & 27 & 12 & 1 & 1\\
   $re1[0.08:0.12]$  & $1\ \!547$ & 733 & 523& 350 & 158 & 44& 12 \\
   $re1[0.12:0.15]$  & $4\ \!348$ & $1\ \!296$ & 717 & 622 & 448 & 146 & 63\\
   $re1[0.15:0.25]$ & $27\ \!774$ & $7\ \!174$ & $1\ \!576$ & 958& 689 & 510 & 303\\
   $re10$  & $48\ \!929$ & $13\ \!170$ & $5\ \!753$ & $3\ \!858$ & $2\ \!466$ &$1\ \!116$ & 605 \\
   $re100$ & $57\ \!068$ & $30\ \!411$ & $21\ \!821$ & $18\ \!933$ & $14\ \!189$ & $6\ \!877$ & $2\ \!787$\\
  \hline
  $In1000N100Me163$ & 0.020 & 0.022 &0.025& 0.030 & 0.035 & 0.040& 0.045\\
  \hline
  $re1[0.005:0.08]$ & 192 & 145 & 97 & 45 & 26 & 14 & 6\\
   $re1[0.08:0.12]$  & 959 & 574 & 352& 223 & 122 & 81& 57 \\
   $re1[0.12:0.15]$  & $2\ \!433$ & $1\ \!675$ & 914 & 389 & 267 & 170 & 107\\
   $re1[0.15:0.25]$ & $21\ \!386$ & $14\ \!733$ & $9\ \!198$ & $4\ \!471$& $2\ \!303$ & $1\ \!388$ & 870 \\
   $re10$  & $41\ \!252$ & $27\ \!075$ & $16\ \!355$ & $6\ \!825$ & $3\ \!981$ & $3\ \!072$ & $1\ \!638$ \\
   $re100$ & $472\ \!422$ & $322\ \!674$ & $200\ \!038$ & $101\ \!111$ & $51\ \!621$ & $28\ \!968$ & $19\ \!351$\\
  \hline
  
\end{tabular}
 \caption{ {Averaged number of additional frequent itemsets after adding variables to the database in the independent case}}
\label{table 11}
\end{table}

\begin{table}[h!]
\centering
\begin{tabular}{|l|c|c|c|c|c|c|c|}
  \hline
   $ARn1000N100Me103$ & 0.015 &0.0175 &0.020& 0.025 & 0.030 & 0.035& 0.040\\
  \hline
   $re1$ & $1\ \!682$ & 882 & 521 & 197 & 92 & 50 & 35\\
   $re10$  & $16\ \!925$ &$8\ \!931$ & $5\ \!310$ & $1\ \!999$ & 920 &502 & 318 \\
   $re100$ & $172\ \!163$ & $91\ \!032$ & $54\ \!143$ & $20\ \!036$ & $9\ \!567$ & $5\ \!195$ & $3\ \!200$\\
  \hline
  $ARn1000N100Me163$ & 0.035 &0.0375 &0.040& 0.045 & 0.050 & 0.055& 0.060\\
  \hline
   $re1$ & $1\ \!183$ & 813 & 607 & 329 & 188 & 126 & 76\\
   $re10$  & $12\ \!627$ &$8\ \!639$ & $6\ \!336$ & $3\ \!388$ & $1\ \!974$ &$1\ \!290$ & 775 \\
   $re100$ & $141\ \!364$ & $91\ \!236$ & $65\ \!233$ & $34\ \!701$ & $20\ \!063$ & $13\ \!400$ & $8\ \!651$\\
  \hline
\end{tabular}
 \caption{ {Averaged number of additional frequent itemsets after adding variables to the database in the AR case}}
\label{table 12}
\end{table}

\begin{figure} [H]
\begin{tikzpicture}[scale = 1]
\begin{scope} [xshift= 1.5 cm]
\begin{axis}[enlarge x limits=false,try min ticks={7},
             legend entries = {PrefRec},
             legend style = {nodes={scale=0.63, transform shape}, at = {(0.52,1)},anchor = north west},  
             height = 4.75cm,
             width = 6.1 cm, 
             axis x line = bottom,
             axis y line = left,
             grid = major,
             title = {$Me103$-$re1$},
             xlabel = {minSup value $\times 10^2$},
             ]
\addplot coordinates {(1.5,0.0104) (1.75,0.0101) (2,0.0099) (2.5,0.0098)  (3,0.0097) (3.5,0.00965) (4,0.00963)};
  
\end{axis}
\end{scope}
\begin{scope}[xshift= 8 cm]
\begin{axis}[enlarge x limits=false,try min ticks={7},
             legend entries = {PrefRec},
             legend style = {nodes={scale=0.63, transform shape}, at = {(0.52,1)},anchor = north west},   
             height = 4.75cm,
             width = 6.1 cm, 
             axis x line = bottom,
             axis y line = left,
             ytick = {0.096,0.098,0.1,0.102},
             yticklabels={0.096,0.098,0.1,0.102},
             grid = major,
             title = {$Me103$-$re10$},
             xlabel = {minSup value $\times 10^2$},
             ]
\addplot coordinates {(1.5,0.1030) (1.75,0.10) (2,0.099) (2.5,0.097)  (3,0.096) (3.5,0.095) (4,0.0945)};
             
\end{axis}
\end{scope}
\end{tikzpicture}

\begin{tikzpicture}[scale = 1]
\begin{scope} [xshift= 0.5 cm]
\begin{axis}[enlarge x limits=false,try min ticks={7},
             legend entries = {PrefRec},
             legend style = {nodes={scale=0.63, transform shape}, at = {(0.52,1)},anchor = north west},  
             height = 4.75 cm,
             width = 6.1 cm, 
             axis x line = bottom,
             axis y line = left,
             grid = major,
             title = {$Me103$-$re100$},
             xlabel = {minSup value $\times 10^2$},
             ]
\addplot coordinates {(1.5,1.035) (1.75,1.007) (2,0.993) (2.5,0.976)  (3,0.966) (3.5,0.958) (4,0.953)};

\end{axis}
\end{scope}
\begin{scope}[xshift= 6.5 cm]
\begin{axis}[enlarge x limits=false,try min ticks={7},
             legend entries = {PrefRec},
             legend style = {nodes={scale=0.63, transform shape}, at = {(0.52,1)},anchor = north west},   
             height = 4.75 cm,
             width = 6.1 cm, 
             axis x line = bottom,
             axis y line = left,
             grid = major,
             title = {$Me163$- $re1$},
             xlabel = {minSup value $\times 10^2$},
             ]
\addplot coordinates {(3.5,0.023) (3.75,0.0222) (4,0.0219) (4.5,0.0217)  (5,0.0216) (5.5,0.0215) (6,0.02145)};

\end{axis}
\end{scope}
\end{tikzpicture}

\begin{tikzpicture}[scale = 1]
\begin{scope} [xshift= 1.5 cm]
\begin{axis}[enlarge x limits=false,try min ticks={7},
             legend entries = {PrefRec},
             legend style = {nodes={scale=0.63, transform shape}, at = {(0.52,1)},anchor = north west},  
             height = 4.75cm,
             width = 6.1 cm, 
             axis x line = bottom,
             axis y line = left,
             ytick = {0.216,0.218,0.220,0.222,0.224},
             yticklabels = {0.216,0.218,0.220,0.222,0.224},
             grid = major,
             title = {$Me163$- $re10$},
             xlabel = {minSup value $\times 10^2$},
             ]
\addplot coordinates {(3.5,0.225) (3.75,0.222) (4,0.221) (4.5,0.218)  (5,0.217) (5.5,0.216) (6,0.215)}; 
  
\end{axis}
\end{scope}
\begin{scope}[xshift= 8 cm]
\begin{axis}[enlarge x limits=false,try min ticks={7},
             legend entries = {PrefRec},
             legend style = {nodes={scale=0.63, transform shape}, at = {(0.52,1)},anchor = north west},   
             height = 4.75cm,
             width = 6.1 cm, 
             axis x line = bottom,
             axis y line = left,
             grid = major,
             title = {$Me163$- $re100$},
             xlabel = {minSup value $\times 10^2$},
             ]
\addplot coordinates {(3.5,2.243) (3.75,2.208) (4,2.193) (4.5,2.170)  (5,2.151) (5.5,2.149) (6,2.148)}; 
             
\end{axis}
\end{scope}
\end{tikzpicture}

\caption{Execution time in seconds to add variables to the database $ARn1000N100Me103$ or $ARn1000N100Me163$}
\label{fig 2.3 simul}
\end{figure}

Table~\ref{table 11} shows that the number of added frequent itemsets increases as the frequency of the added variable increases. In the AR case (Table~\ref{table 12}), the variables having the same distribution law, the addition of 10 variables produces a number of frequent itemsets 10 times greater than the addition of a single variable.  

Likewise, Figures~\ref{fig 2.1 simul},~\ref{fig 2.2 simul}, and~\ref{fig 2.3 simul} show that the execution time depends on the frequency of the added variable, and on the minSup value. It also seems that if the number of added variables is multiplied by 10, then the execution time is also multiplied by 10.
Above all, these figures show that PrefRec is particularly effective for updating the algorithm when a new variable (or more) is taken into account.

\subsubsection{The Moving FIM Application}  

We now present four examples of the Moving FIM application. The first step of the Moving FIM application is the construction of the tree of the frequent itemsets based on the first $q$ frequent items. The second step consists in actualizing the tree: when a new frequent item arrives, the algorithm adds the frequent itemsets that contain this new frequent item, and delete all the frequent itemsets that contain the  most right child of $(\emptyset)$. The number of frequent items therefore remains constant and equal to $q$.
In the experiment, we consider a fixed database. Then, to make the moving FIM work, we process the variables of this database one after the other.

The databases we used are the four previous independent and AR with $n=1000$.

We considered two values of $q$: $100$ and $300$. Moreover, in each experiment, we performed $Q=100$  frequent item replacements.

  The results given in Table~\ref{table 13} and Table~\ref{table 14} concern only the second step of the Moving FIM application, that is, the actualization of the tree. Both tables give the number of frequent itemsets added (freq-add), the number of frequent itemsets deleted (freq-del), and the time necessary for the application. All the values given in these tables are the averages obtained over $50$ simulations.

\begin{table}[h!]
\centering
\begin{tabular}{|l|c|c|c|c|c|c|r|}
  \multicolumn{8}{c}{$In$1000-$N$100-$Me$103}\\
  \hline
   minSup & 0.005 &0.007 &0.01& 0.012 & 0.015 & 0.02& 0.025\\
   \hline
   \multicolumn{8}{|c|}{Moving Fim $q$100}\\
     \hline
     add-freq & $11\ \!422$ & $7\ \!056$& $5\ \!046$ & $3\ \!951$ & $2\ \!697$ & $1\ \!300$ & 640 \\
     del-freq & $11\ \!248$ & $7\ \!070$ & $5\ \!179$ & $4\ \!046$ & $2\ \!859$ & $1\ \!315$ & 735\\
     time & 0.514 & 0.486 & 0.455 & 0.420 & 0.379 & 0.336 & 0.308 \\
  \hline
     \multicolumn{8}{|c|}{Moving Fim $q$300}\\
     \hline
     add-freq & $57\ \!096$ & $26\ \!301$& $15\ \!404$ & $11\ \!929$ & $7\ \!912$ & $3\ \!717$ & $1\ \!742$\\
     del-freq & $55\ \!693$ & $26\ \!100$ & $15\ \!758$ & $12\ \!245$ & $8\ \!407$ & $3\ \!758$ &$2\ \!031$ \\
     time & 1.300 & 1.124 & 0.940 & 0.831 & 0.688 & 0.535 & 0.458 \\
  \hline
    \multicolumn{8}{c}{$In$1000-$N$100-$Me$162}\\
  \hline
   minSup & 0.020 &0.022 &0.025& 0.030 & 0.035 & 0.040 & 0.045\\
   \hline
   \multicolumn{8}{|c|}{Moving Fim $q$100}\\
     \hline
     add-freq & $9\ \!576$ & $7\ \!601$& $5\ \!490$ & $3\ \!753$ & $2\ \!539$ & $1\ \!966$ &$1\ \!502$ \\
     del-freq & $9\ \!334$ & $7\ \!505$ & $5\ \!395$ & $3\ \!644$ & $2\ \!486$ & $1\ \!898$ & $1\ \!537$\\
     time & 0.709 & 0.669 & 0.617 & 0.569 & 0.517 & 0.490 & 0.4595 \\
  \hline
     \multicolumn{8}{|c|}{Moving Fim $q$300}\\
     \hline
     add-freq & $49\ \!157$ & $37\ \!341$ & $24\ \!452$ & $14\ \!363$ & $9\ \!069$ & $6\ \!314$ & $4\ \!408$\\
     del-freq & $47\ \!395$ & $37\ \!307$ & $24\ \!275$ & $13\ \!695$ & $8\ \!738$ & $6\ \!056$ & $4\ \!517$\\
     time & 1.749 & 1.588 & 1.380 & 1.197 & 1.035 & 0.941 & 0.870 \\
  \hline
\end{tabular}
\caption{ {Time in seconds and number of frequent itemsets added and deleted in the Moving Fim application on independent databases}}
\label{table 13}
\end{table}

\vspace*{ 10 cm} 

\begin{table}[h!]
\centering
\begin{tabular}{|l|c|c|c|c|c|c|r|}
  \multicolumn{8}{c}{$ARn$1000-$N$100-$Me$103}\\
  \hline
   minSup & 0.015 &0.0175 &0.020& 0.025 & 0.030 & 0.035& 0.040\\
   \hline
   \multicolumn{8}{|c|}{Moving Fim $q$100}\\
  \hline
  add-freq & $160\ \!904$ & $84\ \!360$ & $49\ \!704$ & $18\ \!882$ &$8\ \!857$ & $4\ \!754$ & $2\ \!999$\\
  del-freq & $163\ \!679$ & $85\ \!787$ & $50\ \!454$ & $19\ \!166$ & $8\ \!976$ & $4\ \!825$ &$3\ \!024$\\
  time   & 0.430 & 0.403 & 0.390 & 0.365 & 0.351 & 0.341 & 0.336\\
  \hline
     \multicolumn{8}{|c|}{Moving Fim $q$300}\\
     \hline
     add-freq & $156\ \!319$ & $81\ \!997$ & $48\ \!448$ & $18\ \!416$ & $8\ \!682$& $4\ \!658$& $2\ \!945$ \\
     del-freq & $163\ \!679$ & $85\ \!787$ & $50\ \!454$ & $19\ \!166$ & $8\ \!976$ & $4\ \!825$ & $3\ \!024$ \\
     time & 0.516 & 0.489 & 0.469 & 0.449 & 0.437 & 0.429 & 0.423 \\
  \hline
    \multicolumn{8}{c}{$ARn$1000-$N$100-$Me$163}\\
      \hline
      minSup & 0.0350 &0.0375 &0.040 & 0.045 & 0.050 & 0.055& 0.060\\
      \hline
   \multicolumn{8}{|c|}{Moving Fim $q$100}\\
     \hline
     add-freq & $130\ \!831$ & $90\ \!045$ & $65\ \!222$ & $34\ \!839$ & $20\ \!341$& $13\ \!144$ & $7\ \!926$\\
     del-freq & $128\ \!819$ & $88\ \!402$ & $64\ \!321$ & $34\ \!350$  & $20\ \!022$ & $12\ \!988$ & $7\ \!827$\\
     time & 0.523 & 0.498 & 0.481 & 0.457 & 0.443 & 0.429 & 0.420 \\
  \hline
     \multicolumn{8}{|c|}{Moving Fim $q$300}\\
     \hline
     add-freq & $132\ \!984$ & $91\ \!861$& $66\ \!218$ & $35\ \!414$ & $20\ \!742$ & $13\ \!363$ & $8\ \!051$\\
     del-freq & $128\ \!819$ & $88\ \!402$ & $64\ \!321$ & $34\ \!349$ & $20\ \!022$ & $12\ \!987$ & $7\ \!827$ \\
     time & 0.674 & 0.655 & 0.638 & 0.612 & 0.595 & 0.58 & 0.575 \\
  \hline
\end{tabular}
\caption{ {Time in seconds and number of frequent itemsets added and deleted in the Moving Fim application on AR databases}}
\label{table 14}
\end{table} 

Table~\ref{table 13} and Table~\ref{table 14} show that the variations associated with adding and removing a variable  depend mainly on the values of minSup and $q$. Moreover, the averaged number of frequent itemsets added and deleted is much higher in the case of dependent database than in the case of independent database. But, for both types of databases, there is a stability in the number of frequent itemsets during this application, the number of added frequent itemsets being always close to that of deleted frequent itemsets.

PrefRec is very performant for such an actualization of the tree of the frequent itemsets. This is due to the nature of the underlying tree.

\section{Conclusion}

\label{section conclusion}

In this paper we  presented PrefRec, a new algorithm for fast mining of frequent itemsets and discovering all significant association rules  in large datasets.  The logic of the algorithm is based on a function which, starting from a tree of frequent itemsets, allows to find the new tree of frequent itemsets when a new item is added to the base.  The main property of the algorithm is its recursiveness with respect to the sequence of variables. This makes it very useful when the processed dataset is not fixed and can evolve by adding new variables. Another important property of the algorithm  is that it gives the list of frequent itemsets in a very useful tree structure. In particular this makes it possible to remove from the list of frequent itemsets, in a very simple and very fast way, the first item of the base and all the frequent itemsets which contain it. This makes PrefRec a very efficient algorithm in the case of a moving base, i.e. a base where the sequence of variables changes by removing the first and adding a new one at the end of the sequence. 
A large set of experiments have been carried out to compare the performance of PrefRec and that of Eclat/LCM, Fp-Growth and Apriori and to evaluate the recursion properties of PrefRec. In the case of a fixed base, these experiments show that PrefRec supports the comparison well. In the case of a non fixed base it is much more efficient.

\paragraph{Acknowledgment} We would like to thank the Maisons du Monde data team for the numerous discussions we had on the various applications of the PrefRec algorithm.

\end{document}